\documentclass[12pt]{article}

\usepackage{amssymb,amsmath}

\newenvironment{eq}[1]
{\[\begin{array}{#1}}{\end{array}\]}

\let\rvec=\vec        
\let\Sup=\sup        

\def\intq3{\int{d^3q\over q_0(2\pi)^3}}

\def\plintaqq3{{}^\pl\hskip-2.5mm\int{d^3q\over 2|\rvec q|(2\pi)^3}}

 \def\20strich{ \rule[.5mm]{20mm}{.5mm}  }
 \def\40strich{\rule[.5mm]{40mm}{.5mm}}
 \def\12strich{\rule[.5mm]{12mm}{.5mm}}

\def\plint {\hskip-2mm{^\pl\hskip-3mm\int}}

\def\lq{\hskip-0.5mm\setminus\hskip-0.5mm} 
  
 \def\({\Bigl(}
\def\){\Bigr)}   
 \def\|{\Big|}
\def\then{\Rightarrow}  
 \def\o{\circ}
\def\m{\bullet}    
\def\x{\times}
   
\def\ox{\otimes}

\def\pl{{~\oplus~}}

\def\PL{\displaystyle \bigoplus}

\def\mid{\big\bracevert}

\def\sub{\subseteq}
\def\subnoteq{\subset}
\def\Sup{\supseteq}
\def\supnoteq{\supset}

\def\and{\wedge}

\def\AND{\displaystyle\bigwedge}
\def\od{\vee}

\def\OD{\displaystyle\bigvee}

\def\rin{{\,\in\kern-.42em\in}}
 
 \def\diag{{\,{\rm diag}\,}}

\def\spec{\,{\rm spec}\,}

\def\tr{{\,{\rm tr }\,}}
\def\det{\,{\rm det }\,}
\def\id{\,{\rm id}}

\def\sx{~\rvec\x~\!}

\def\irrep{{{\bf irrep\,}}}

\def\C{\mathbb C}

\def\N{\mathbb N}
\def\O{\mathbb O}

\def\R{\mathbb R}

\def\Z{\mathbb Z}

\def\p{\partial}
 
\def\al{\alpha}  \def\be{\beta} \def\ga{\gamma}
\def\de{\delta}  \def\ep{\epsilon}  
\def\th{\theta}   \def\vth{\vartheta} \def\io{\iota}
\def\ka{\kappa}   \def\la{\lambda}   \def\si{\sigma}
   \def\om{\omega} \def\Om{\Omega}
\def\phi{\varphi} 
   
    \def\La{\Lambda}

\def\vec#1{\underline{\bf vec}_{#1}}

\def\GL{{\bf GL}}  
\def\SL{{\bf SL}}
\def\U{{\bf U}} 
 
\def\O{{\bf O}}   
\def\SU{{\bf SU}} 
 
\def\SO{{\bf SO}}

\def\D{{\bl D}}

\def\norm#1{\parallel\hskip-.3mm #1 \hskip-.3mm\parallel}
\def\d#1{{\check{#1}}}
\def\lrangle#1{\langle#1\rangle}

\def\rstate#1{|#1\rangle}
\def\lstate#1{\langle#1|}

\def\brack#1{\lbrack#1\rbrack}
\def\Brack#1{\Bigl\lbrack#1\Bigr\rbrack}

\def\ro#1{{\rm #1}}
\def\bl#1{{\bf {#1}}}
\def\cl#1{{\cal #1}}

\def\ol#1{\overline{#1}}

\def\sprod#1#2{\langle#1|#2\rangle}

\def\com#1#2{\lbrack#1,#2\rbrack}

\def\acom#1#2{\{#1,#2\}}

\def\map{\longrightarrow}

\def\lrmap{\leftrightarrow}

\def\dmap{\Big\downarrow}

\def\mape{\longmapsto}

\def\Diagr#1#2#3#4#5#6#7#8{\begin{matrix}
\noalign{\vskip5mm}
      \hskip-4mm&              &{\scriptstyle #5}&              &     \cr
\noalign{\vskip-2mm}
      \hskip-4mm& #1           & \map           & #2           &     \cr
{\scriptstyle #8}\hskip-3mm   &\dmap         &    &\dmap&\hskip-4mm{\scriptstyle#6} \cr
      \hskip-4mm& #4           & \map           & #3           &     \cr
      \noalign{\vskip-2mm}
      \hskip-4mm&              &{\scriptstyle#7}&              &     \cr
\noalign{\vskip5mm} 
            \end{matrix}}

\advance\topmargin by -1.6cm
 
\begin{document}

\begin{titlepage} 
$~$
\vskip5mm
\hfill MPP-2006-11
\vskip25mm

\centerline{\bf RELATIVITIES AND HOMOGENEOUS SPACES I}
\vskip5mm
\centerline{\bf --Finite dimensional  relativity representations--}
\vskip15mm
\centerline{
Heinrich Saller\footnote{\scriptsize hns@mppmu.mpg.de} }
\centerline{Max-Planck-Institut f\"ur Physik}
\centerline{Werner-Heisenberg-Institut}
\centerline{M\"unchen, Germany}

\vskip25mm

\centerline{\bf Abstract}
\vskip15mm
Special relativity, the symmetry breakdown in the electroweak standard model,
 and
the dichotomy of the spacetime related trans\-for\-ma\-tions  
with the Lorentz
group,  on the one side, and the chargelike trans\-for\-ma\-tions 
with the hypercharge and isospin
group, on the other side,  are discussed under the common concept of ``relativity". A relativity is
defined by classes $G/H$ of a ``little" group in a 
``general" group of operations.
Relativities are representable as linear trans\-for\-ma\-tions that are considered for
five physically relevant examples.

\end{titlepage}

\newpage
{\small\tableofcontents}

\newpage

\section{Five relativities for an introduction }

Basic physical theories
involve both external and internal  degrees of freedom
 that are acted on, respectively,  by operations
from the  Poincar\'e group, i.e.,  Lorentz group and spacetime translations, 
and by   operations from the hypercharge,
 isospin and color group. 
The properties of all basic interactions and particles 
are determined and characterized by invariants and
eigenvalues for these operation groups. Although 
the product of external and internal operations 
in the acting group is  direct, 
the internal ``chargelike" operations are coupled to
the  external ``spacetimelike" ones: 
any spacetime translation is accompanied by a chargelike operation. 
This 
is implemented by the gauge fields in the standard
 model of electroweak and strong interactions.  In the following,
the dichotomy and the connection of external and internal operations will be 
discussed
 under the label ``unitary relativity", especially with respect to its
 re\-pre\-sen\-ta\-tions by interactions and particles.

To see its general and its specific structures, unitary relativity will 
 be introduced and considered  as one example in five  relativities: 
perpendicular relativity as realized after discovering the 
 surface of the earth to be spherical,
 rotation relativity, or space and time relativity, 
  as used in what we call special relativity
 with ``timelike" and ``spacelike" translations,
Lorentz group relativity, or Minkowski spacetime relativity, as an important ingredient of  general
relativity, electromagnetic relativity as formulated in the standard model 
of electroweak
interactions \cite{WEIN} and finally, and that is mostly  new, unitary relativity.

Relativity will be defined by operation groups, an example:
In special relativity, the distinction of your rest system determines a
decomposition of spacetime translations into 
time and position translations. Compatible with this decomposition
is your position rotation group $\SO(3)$ as a subgroup of the orthochronous
Lorentz group $\SO_0(1,3)$. There are as many decompositions of
 spacetime into time and position
as there are rotation groups in a Lorentz group.
The rotation group  classes  are pa\-ra\-me\-tri\-zable by the points of
a one shell 3-di\-men\-sio\-nal hy\-per\-bo\-loid $\cl Y^3\cong \SO_0(1,3)/\SO(3)$
that give the mo\-men\-ta (velocities) for all
possible motions.
Another example: The perpendicularities of mankind, if earthbound, are characterized  by the 
axial rotation groups in a rotation group and  pa\-ra\-me\-tri\-zable by 
coordinates of the earth's 
surface $\Om^2\cong\SO(3)/\SO(2)$.

Now in general: The choice of an ``idolized" operation group $H$
in a ``general" operation group $G$ picks one element in the $G$-symmetric
space  $G/H$, which stands for  the relativity of the ``idolized" group,
called $H$-relativity. 
An ``idolization" \cite{BAC} goes, negatively, 
with the ``narrow-minded" assumption of an absolute
point of view or, positively, with the distinction of a 
smaller operation symmetry,  enforced, e.g.,  
 by initial or boundary conditions. 
Important examples are degenerate ground states (``spontaneous symmetry
breakdown") where an ``interaction-symmetry"  $G$ is reduced to 
a ``particle-symmetry" $H$, e.g., the degenerate ground states of
superconductivity, of superfluidity, of a ferromagnetic or of the electroweak
standard model. The ground state degeneracy
is characterized by the symmetric space $G/H$.

This gives the first four columns of the following table,
which together with the last one will be discussed with their re\-pre\-sen\-ta\-tions
in more detail below
\vskip2mm
{\scriptsize
\begin{eq}{c}

\hskip-10mm
\begin{array}{|c||c|c|c|c|}\hline

\hbox{relativity}&
\begin{array}{c}
\hbox{``general"}\cr
\hbox{group }G\cr
(r,r_\cl R)\cr\end{array}
&\begin{array}{c}
\hbox{``idolized"}\cr
\hbox{subgroup }H\cr
\end{array}&
\begin{array}{c}
\hbox{homogeneous space }G/H\cr
\end{array}
&\begin{array}{c}\hbox{relativity}\cr\hbox{parameters}\end{array}\cr
\hline\hline

\begin{array}{c}\hbox{axial rotation}\cr \hbox{(perpendicular)}\cr
\hbox{relativity}\cr\end{array}&
\begin{array}{c}\SO(3)\cr\sim\SU(2)\cr (1,0)\end{array}&
\begin{array}{c}\SO(2)\cr
\end{array}&
\begin{array}{c}\hbox{2-sphere}\cr\Om^2\cong\SO(3)/\SO(2)\cr
\hfill\cong\SU(2)/\SO(2)
\end{array}&
\begin{array}{c}
\hbox{2 transversal}\cr
\hbox{coordinates}\cr
\end{array}\cr 
\hline

\begin{array}{c} \hbox{rotation}\cr\hbox{(special)}\cr
\hbox{relativity}\cr\end{array} &
\begin{array}{c}\SO_0(1,3)\cr\sim\SL(\C^2)\cr (2,1)\end{array}&
\begin{array}{c}\hfill\SO(3)\cr\sim\SU(2)\cr \end{array}&
\begin{array}{c}\hbox{3-hy\-per\-bo\-loid}\cr
\cl Y^3\cong\SO_0(1,3)/\SO(3)\cr \hfill\cong\SL(\C^2)/\SU(2)\cr
\end{array}&
\begin{array}{c}
\hbox{3 mo\-men\-ta}\cr
\end{array}\cr 
  \hline 
  
\begin{array}{c} \hbox{Lorentz group}\cr
\hbox{(general)}\cr\hbox{relativity}\cr\end{array} &
\begin{array}{c}\GL(\R^4)\cr (4,4)\cr\end{array}&
\begin{array}{c}\O(1,3)\cr\end{array}&
\begin{array}{c}\hbox{tetrad or metric manifold}\cr
\cl M^{10}\cong\GL(\R^4)/\O(1,3)\cr
\hfill\cong\D(1)\x\SO_0(3,3)/\SO_0(1,3)\cr
\end{array}&
\begin{array}{c}
\hbox{10 components}\cr
\hbox{for metric tensor}\cr
\end{array}\cr  \hline

\begin{array}{c}\hbox{electro-}\cr\hbox{magnetic}\cr\hbox{relativity}\end{array}&
\begin{array}{c}\U(2)\cr(2,0)\end{array}&
\begin{array}{c}\U(1)_+\cr\end{array}&
\begin{array}{c}\hbox{Goldstone manifold}\cr\cl G^3\cong\U(2)/\U(1)_+\cr
\end{array}&
\begin{array}{c}
\hbox{3 weak}\cr\hbox{coordinates}\end{array}\cr\hline

\begin{array}{c}\hbox{unitary}\cr\hbox{relativity}\cr\end{array}&
\begin{array}{c}\GL(\C^2)\cr (4,2)\end{array}&
\begin{array}{c}\U(2)\cr \end{array}&
\begin{array}{c}\hbox{positive 4-cone}\cr\cl D^4\cong\GL(\C^2)/\U(2)\cr
\hfill\cong\D(1)\x\SO_0(1,3)/\SO(3)\cr
\end{array}&
\begin{array}{c}
\hbox{4 spacetime}\cr
\hbox{coordinates}\end{array}\cr 
\hline

\end{array}\cr
\cr
\hbox{\bf orientation manifolds of five
relativities}

\end{eq}}

Somewhat in accordance with the historical development,
the ``general" operations of one relativity can constitute the ``idolized" 
group
of the next relativity as seen in the two chains ending in  full general
linear groups, a real  one for spacetime concepts,
 from flat to spherical earth to special and 
general
relativity, and a complex one for interactions, from electromagnetic to electroweak 
trans\-for\-ma\-tions and their spacetime (gauge) dependence: 
\begin{eq}{llll}
\SO(2)&\subnoteq\SO(3)&\subnoteq\SO_0(1,3)&\subnoteq\GL(\R^4),\cr
\U(1)_+&\subnoteq\U(2)&\subnoteq\GL(\C^2).
\end{eq}

All groups in the five relativities considered  are
 real Lie groups.
All ``general" groups are reductive, for perpendicular and rotation
relativity even semi\-simple.
Perpendicular and electromagnetic relativity have a compact ``general" group.
With the exception of Lorentz group relativity, all ``idolized"
 groups are compact subgroups.
The 2nd column contains  the dimension  of 
the maximal abelian subgroups, 
which is the rank $r$ of the group $G$ generating Lie algebra
$L=\log G$,  
and  of  the maximal noncompact abelian subgroups,
i.e., the real rank $r_\cl R$. 
With the exception of $\GL(\R^4)$, the maximal abelian
subgroups allow a unique decomposition 
into compact Cartan torus and noncompact  Cartan plane.
A Cartan torus is a direct product of circle groups in the form  $\U(1)=\exp
i\R$ or $\SO(2)\cong\exp \si_3 i\R$, a Cartan plane 
is a direct product of 
 additive line groups $\R$ (translations), which also can be used 
  in the multiplicative form  $\D(1)=\exp
\R$ or $\SO_0(1,1)\cong\exp \si_3\R$.
The  rank gives the number of 
independent invariants, rational or continuous $r=n_\cl I+n_\cl R$,
that characterize a $G$-re\-pre\-sen\-ta\-tion.
The real rank is the maximal
number  of the continuous invariants $n_\cl R\le r_\cl R$.

Unitary relativity $\GL(\C^2)/\U(2)\cong\D(1)\x\SO_0(1,3)/\SO(3)$, 
i.e.,  the complex linear relativization of the 
maximal compact subgroup with the internal ``chargelike"
hypercharge and isospin operations $\U(2)$, 
is pa\-ra\-me\-tri\-zed by a 
noncompact real 4-di\-men\-sio\-nal 
 homogeneous space, called causal spacetime $\cl D^4$,
 a name to be justified below.
Unitary relativity is visible in the spacetime dependence of quantum fields,
which represent  the internal operations.
The re\-pre\-sen\-ta\-tions of unitary relativity 
$\cl D^4$ with a 2-di\-men\-sio\-nal Cartan plane are characterized by
two continuous invariants, which, in appropriate units, can be taken as two
masses. The $\cl D^4$-re\-pre\-sen\-ta\-tions  determine the 
 spacetime interactions with their normalization, especially the gauge
 interactions with  their coupling constants related to the ratio of the two invariants,
and, for the $\cl D^4$-tangent translations $\R^4$,  the particles
and their masses. The common language for interactions 
and elementary particles 
is the re\-pre\-sen\-ta\-tion theory and harmonic analysis 
of unitary relativity (more below).

There is a mathematical framework, almost tailored for relativities:
the theory of induced re\-pre\-sen\-ta\-tions,
pioneered by Frobenius \cite{FRO}, used  for free
particles by Wigner \cite{WIG} and  worked out for noncompact groups especially by Mackey \cite{MACK1}.
There, a subgroup $H$-re\-pre\-sen\-ta\-tion induces
a full group $G$-re\-pre\-sen\-ta\-tion leading
to a $G\x H$-re\-pre\-sen\-ta\-tion  as subre\-pre\-sen\-ta\-tion of the two-sided regular
$G\x G$-re\-pre\-sen\-ta\-tion. 
Such a dichotomic trans\-for\-ma\-tion property with a doubled group,
$G\x G$ as group and ``isogroup", is
familiar, with respect to the Lorentz and the isospin group,
$\SU(2)\x\SU(2)$ as spin and isospin, 
from the fields in the electroweak standard model.
Especially for noncompact nonabelian groups, the theory
is not easy to penetrate. All the mathematical details are given in the 
textbooks of Helgason \cite{HELG2}, Knapp \cite{KNAPP} and Folland \cite{FOL}
and, especially for distributions, of Treves \cite{TREV}. 

In the following, only some motivating and qualititive mathematical 
remarks will be given
with respect to this theory, which will be  used in physical implementations.
The first part of this paper  works with 
 finite-di\-men\-sio\-nal relativity structures,
which may be not so familiar in such a conceptual framework.
After a pa\-ra\-me\-tri\-zation of the relativity manifold $G/H$, there will be given its
fundamental representations, called transmutators, which mediate
the transition from an idolized group $H$ to the full group $G$. 
With the fundamental transmutators all finite dimensional
relativity representations can be constructed as used, e.g., in the
transition from the fields for the electroweak interactions to the 
asymptotic particles.

The mathematically more demanding second part (``Spacetime and unitary
relativity") uses re\-pre\-sen\-ta\-tions of noncompact
operation groups on Hilbert spaces, necessarily infinite-di\-men\-sio\-nal for
faithful re\-pre\-sen\-ta\-tions.

\section{Relativity parameters}

There are operation-induced  parameters 
for the real homogeneous relativity spaces
$G/H$, e.g., the three mo\-men\-ta (velocities) for rotation (special) relativity
or the  three weak coordinates of electromagnetic relativity as used in the
mass modes of the   three weak bosons. 

The action of a ``general" group $G$ on a set $S$, denoted by $\m$,
 decomposes $S$ into disjoint orbits
$G\m x$ for $x\in S$ that are isomorphic to subgroup classes $G\m x\cong G/H$
where the ``idolized" group $H$ is the fixgroup (fixer, ``little" group,
isotropy group) $G_x$ 
of the $G$-action.
The elements of homogeneous spaces $gH\in G/H$
are group subsets (cosets),
e.g., position rotation groups in a Lorentz group
or electromagnetic trans\-for\-ma\-tion groups in a hypercharge-isospin group. The
cosets have re\-pre\-sen\-tatives
$g_r\in gH\in G/H$, written as $g_r\rin G/H$, which can be characterized by what will be called ``relativity
parameters", a real pa\-ra\-me\-tri\-zation of the subgroup classes. 

Relativity parameters can be obtained via orbit pa\-ra\-me\-tri\-zations.
The real Lie groups considered are linear 
groups $H\sub G\sub \GL(V)$,  acting on 
real or complex vector spaces $V$.
The orbit $G\m x$ pa\-ra\-me\-tri\-zes the homogeneous space $G/H$
by $V$-vectors and their components with respect to a basis,
\begin{eq}{rl}
x\in V,~ H&\cong G_x=\{g\in G\mid g\m x=x\}\cr
\then G/H&\cong G\m x\sub V.
\end{eq}

\subsection{Weak  coordinates for electromagnetic relativity}

The hypercharge-isospin group $\U(2)$ 
acts, in the defining re\-pre\-sen\-ta\-tion, on 
a complex 2-di\-men\-sio\-nal vector space: 
\begin{eq}{l}
\U(2)\ni u=
e^{i\al_0}
{\scriptsize\begin{pmatrix}
e^{i\al_3}\cos{\th\over2}&-e^{-i\phi}\sin{\th\over2}\cr
e^{i\phi}\sin{\th\over2}&e^{-i\al_3}\cos{\th\over2}\cr\end{pmatrix}}.
\end{eq}Each nontrivial vector has  a $\U(1)$-isomorphic fixgroup, e.g. 
$e^2$, which defines $\U(1)_+$ as an ``idolized" electromagnetic subgroup, 
\begin{eq}{l}
\C^2\cong V\ni e^2={\scriptsize\begin{pmatrix}0\cr1\cr\end{pmatrix}}\then
{\scriptsize\begin{pmatrix}
e^{2i\al_0}&0\cr 0&1\cr\end{pmatrix}}\in\U(2)_{e^2}=\U(1)_+.\cr
\end{eq}The orbits of the chosen vector, here $u\m e^2$, and 
its $\U(2)$-orthonormal partner, here $(u\m e^2)_\bot$, 
give the two columns of the
matrix pa\-ra\-me\-tri\-zation $v\in\U(2)$ of the Goldstone manifold $\cl G^3$,  
\begin{eq}{rl}

\cl G^3\cong\U(2)/\U(1)_+&\cong \{((u\m e^2)_\bot,u\m e^2)=v\mid
u\in\U(2)\},\cr
v&
={\scriptsize\begin{pmatrix}
e^{i(\al_3-\al_0)}\cos{\th\over2}&- e^{-i(\phi-\al_0)}\sin{\th\over2}\cr
e^{i(\phi-\al_0)}\sin{\th\over2}&e^{-i(\al_3-\al_0)}\cos{\th\over2}\cr\end{pmatrix}}.\cr

  \end{eq}In the standard model of electroweak interactions
 the vector space $V\cong\C^2$ desribes the chargelike degrees of freedom of the
 Higgs field.
 The three weak  parameters
  $(\al_3-\al_0,  \phi-\al_0,\th)$
pa\-ra\-me\-tri\-ze  electromagnetic relativity.
As manifold, not as group, 
the Goldstone
manifold $\cl G^3$ is isomorphic to $\SU(2)$.

\subsection{Orbits of metric tensors}

With the exception of electromagnetic relativity, all
relativity parameters will by given by the ``general" group $G$-orbit
of a metric invariant under the action of an ``idolized" subgroup $H$. 
In this context, the homogeneous space $G/H$ for $H$-relativity
was called,  by Weyl \cite{WEYLRZM}, orientation manifold of the metric 
(bilinear or sesquilinear product).

The invariance of a metric\footnote{\scriptsize
 Usual notation 
 for the metric $\ga^{\mu\nu}=g^{\mu\nu}$, here $g\in G$ is reserved for group
 elements.}
 $\ga$ with respect to the action of
a linear group,
\begin{eq}{l}
\GL(V)\supnoteq H\ni g,~~ \ga(x,y)\mape \ga(g\m x,g\m y)=\ga(x,y)\hbox{ for all
}x,y\in V,
\end{eq}gives the pa\-ra\-me\-tri\-zation
of the fixgroup classes by
 the orbit of the metric tensor $\ga$, 
\begin{eq}{l}
 \{g\in G\sub \GL(V) \mid g\o \ga\o g^*=\ga\}=H\then  
\{g\o \ga\o g^*\mid g\in G\}\cong G/H.
\end{eq}

\subsection{Metric tensor for Lorentz group relativity}

A bilinear form (metric) of a vector space $V$ is a power two
tensor $\ga\in V^T\ox V^T$ with the dual vector space $V^T$ (linear forms).
For an $n$-di\-men\-sio\-nal space, the subspaces 
$V^T\and  V^T$, totally antisymmetric, denoted by $\and$, 
 and $V^T\od V^T$, totally symmetric,  denoted by 
 $\od$, have the dimensions 
${n\choose 2}$ and ${n+1\choose 2}$ respectively.

A real vector space $V\cong\R^n$ has a causality structure by
embedding  the cone of the positive numbers $\R_+\map V_+\subnoteq V$ 
into the ``future" cone of the vector space
$x\succeq 0\iff x\in V_+$. A nontrivial ``future"
cone $V_+\ne\{0\}$ can be defined by 
a bilinear symmetric
  form    with  ``causal" signature $(t,s)=(1,s)$
  invariant under the
  generalized Lorentz group $\SO_0(1,s)$. 
 Such a causality structure for $V\cong\R^{1+s}$
 is familiar for time $\R$ with total order and
 Minkowski spacetime $\R^4$ with the special relativistic 
 partial order.

Any metric tensor
of $V\cong \R^4$ with causal signature $(1,3)$, 
e.g.,   an orthonormal Lorentz metric tensor,
\begin{eq}{l}
V\cong\R^4,~~\eta
={\scriptsize\begin{pmatrix}1&0\cr0&-\bl1_3\cr\end{pmatrix}}\in V^T\od V^T,\cr 
\end{eq}defines
an ``idolized" Lorentz group as invariance goup, 
Its $\GL(\R^4)$-orbit leads  to a pa\-ra\-me\-tri\-zation of  
 the metric manifold 
 with dimension ${5\choose2}=10$  
for Lorentz group relativity
\begin{eq}{rl}
\cl M^{10}\cong \GL(\R^4)/\O(1,3)&\cong
\{h\o\eta\o h^T= \ga \mid h\in\GL(\R^4)\},\cr
 \ga&\cong \ga^{\mu\nu}=h^\mu_j\eta^{jk}h_j^\nu=\ga^{\nu\mu};
\end{eq}here $\mu,j\in\{0,1,2,3\}$.

\subsection{Spherical coordinates for perpendicular relativity}

With the local isomorphy of the rotation group to the spin group $\SO(3)\sim\SU(2)$
an ``idolized" axial rotation subgroup $\SO(2)\subnoteq \SU(2)$ is given by the invariance group of
the hermitian and traceless  
Pauli matrix $\si_3$. Its $\SU(2)$-orbit leads to the 
2-sphere pa\-ra\-me\-tri\-zation of perpendicular relativity,
\begin{eq}{rl}
\si_3={\scriptsize\begin{pmatrix}1&0\cr0&-1\cr\end{pmatrix}},~~ 
\Om^2\cong\SO(3)/\SO(2)&\cong\{
u\o\si_3\o u^\star={\rvec x\over r}
\mid u\in\SU(2)\},\cr
{\rvec x\over r}&
\cong{\rvec x^\al_\be\over r}=u^\al_j\si_3{}^j_ku^\star{}_\be^k;
\end{eq}here $\al,j\in\{1,2\}$. The two angles 
(spherical coordinates) in the traceless hermitian matrix ${\rvec x\over r}$
can be pa\-ra\-me\-tri\-zed by 
three position translations with one condition
for the determinant, 
\begin{eq}{l}
{\rvec x\over r}={\rvec x^\star\over r}=
{\scriptsize\begin{pmatrix}\cos \th&e^{-i\phi}\sin\th\cr
e^{i\phi}\sin \th& -\cos \th\cr\end{pmatrix}}=
{1\over r}
{\scriptsize\begin{pmatrix}x_3&x_1-ix_2\cr x_1+ix_2& -x_3\cr\end{pmatrix}}\cr
\hbox{with }\tr {\rvec x\over r}=0\hbox{ and }-\det{\rvec x\over r}=
{\rvec x^2\over r^2}=1. 
\end{eq}The restriction uses the
rotation $\SO(3)$-invariant product $\rvec x^2=x_3^2+x_1^2+x_2^2$ in three dimensions.

\subsection{Momenta for rotation relativity}

An ``idolized" rotation group $\SO(3)$ 
in a Lorentz group $\SO_0(1,3)$ is characterized by a distinguished
definite metric of a real 3-di\-men\-sio\-nal vector space
(position), e.g.,  $\ga=\bl 1_3$. 
Similarly, one can work with a sesquilinear
scalar product $\de$ of a complex 2-di\-men\-sio\-nal 
space $V\cong\C^2$ invariant under the locally isomorphic spin group
$\SU(2)\sim\SO(3)$ in the special linear group $\SL(\C^2)\sim\SO_0(1,3)$.
The $\SL(\C^2)$-orbit of the metric pa\-ra\-me\-tri\-zes rotation relativity
by the points of an energy-mo\-men\-tum 3-hyperboloid,
\begin{eq}{rl}
\de=\bl1_2,~~\cl Y^3\cong\SO_0(1,3)/\SO(3)&\cong
\{s\o\de\o s^\star={ q\over m}
\mid s\in\SL(\C^2)\},\cr
{ q\over m}&\cong
{ q^A_{\dot B}\over m}=s^A_\al\de^\al_\be s^\star{}_{\dot B}^\be;
\end{eq}here $A,\al\in\{1,2\}$. The three real 
hyperbolic coordinates in the hermitian
 matrix $ {q\over m}$
can be chosen from  four en\-er\-gy-mo\-men\-ta with one condition
for the  determinant, 
\begin{eq}{l}
{ q\over m}={ q^\star\over m}=
{\scriptsize\begin{pmatrix}\cosh2\be+ \cos\th\sinh2\be& e^{-i\phi}\sin\th\sinh2\be\cr
e^{i\phi}\sin \th\sinh2\be& \cosh2\be- \cos\th\sinh2\be\cr\end{pmatrix}}=
{1\over m}
{\scriptsize\begin{pmatrix}q_0+q_3&q_1-iq_2\cr q_1+iq_2& q_0-q_3\cr\end{pmatrix}}\cr
\hbox{with }\det  {q\over m}={q^2\over m^2}=1.
\end{eq}The restriction of the four en\-er\-gy-mo\-men\-ta
to the three mo\-men\-ta uses the $\SO_0(1,3)$-invariant 
bilinear form $q^2=q_0^2-\rvec q^2$.

\subsection{Spacetime future  for unitary relativity}

An ``idolized" unitary group $\U(2)$, 
called hyperisospin group, a maximal compact subgroup
of the general linear group $\GL(\C^2)$, called
extended Lorentz group,
is given by the invariance group of a scalar product $\de$
of a complex 2-di\-men\-sio\-nal vector space. The $\GL(\C^2)$-orbit  defines
  four real parameters for unitary relativity, i.e.,  
for the orientation  manifold of the $\U(2)$-scalar product, 
\begin{eq}{rl}
\de=\bl1_2,~~\cl D^4\cong\GL(\C^2)/\U(2)&\cong\{\psi\o\de\o\psi^\star= x\mid
\psi\in\GL(\C^2)\},\cr
 x&\cong x^A_{\dot B}=\psi^A_\al\de^\al_\be\psi^\star_{\dot B}{}^\be, \cr
 x&= x^\star=
{\scriptsize\begin{pmatrix}x_0+x_3&x_1-ix_2\cr x_1+ix_2&
x_0-x_3\cr\end{pmatrix}};\cr
\end{eq}here $A,\al\in\{1,2\}$.
These four real orbit  parameters
characterize  the strictly positive elements in the $C^*$-algebra of complex 
$(2\x 2)$ matrices,
\begin{eq}{rl}
x=\psi\o\psi^\star&\iff
x=x^\star\hbox{ and }\spec  x>0\cr &\iff \det x=x^2>0\hbox{ and }\tr x=2x_0>0.
\end{eq}They describe the absolute modulus set
in the polar decomposition of $\GL(\C^2)$
into noncompact classes for the maximal compact 
group with the unitary phases, 
\begin{eq}{l}
\GL(\C^2)\ni \psi=|\psi|\o u\in\D(2)\o\U(2),\cr
\GL(\C^2)/\U(2)\cong\D(2)\ni |\psi|= \sqrt{\psi\o \psi^\star}=\sqrt x.
\end{eq} 

The positive matrices $ x$ 
are pa\-ra\-me\-tri\-zable by the 
points of the open future cone in
flat Minkowski spacetime,
\begin{eq}{l}
\cl D^4\cong\R^4_+=\{x\in \R^4\mid x^2>0, x_0>0\}.
\end{eq}The cone manifold is embeddable into its own tangent space,
the spacetime translations $\R^4\supnoteq\cl D^4$.
They inherit the  action of the dilation extended orthochronous Lorentz group
$\GL(\C^2)/\U(1)\cong\D(1)\x\SO_0(1,3)$,
which constitutes the homogeneous part in
 the extended Poincar\'e group $[\D(1)\x\SO_0(1,3)]\sx\R^4$.


\section{Relativity transitions}

Elements of a relativity, i.e., of a homogeneous space  $G/H$,
are related to each other by the action of
the full group $G$, e.g., different perpendicularities
by rotations of the earth's surface or 
different nonrelativistic space-times by Lorentz trans\-for\-ma\-tions
of spacetime.

With  real parameters for $H$-relativity  $G/H$ 
one can partly pa\-ra\-me\-tri\-ze the
``general" group $G$. 
Each coset can be given a defining re\-pre\-sen\-tative $g_r\in gH\sub G$
as linear operation on a complex vector space.
Such re\-pre\-sen\-tatives 
have a characteristic two-sided  $G\x H$  trans\-for\-ma\-tion
behavior in the group $G\x G$, called relativity transition or transmutation
from the ``general" group to the ``idolized" group:
A left multiplication  of the re\-pre\-sen\-tative $g_r\in gH$ by $k\in G$
hits the chosen re\-pre\-sen\-tative $(kg)_r\in kgH$ up to a right multiplication
with an $H$-element,
\begin{eq}{l}
k\in G,~~ kg_r=(kg)_r h(g_r,k)
\hbox{ with }h(g_r,k)\in H.
\end{eq}The  group action $k\in G$ is accompanied by 
an action from  the ``idolized" subgroup  $h(g_r,k)\in H$, which depends on the
re\-pre\-sen\-tative $g_r$.
It is called Wigner
element and Wigner subgroup-operation, in generalization of the familiar Wigner rotation, which arises
from a Lorentz trans\-for\-ma\-tion of a boost.

\subsection{From interaction group  to particle group}

An example where both electromagnetic relativity
with the transition $\U(2)\to\U(1)_+$ 
and rotation (special) relativity 
with the transition $\SL(\C^2)\to\SU(2)$ 
play a role is the transition from relativistic electroweak interaction fields 
to particles in the standard model \cite{S003},
\begin{eq}{c}
\begin{array}{ccccccc}
\SL(\C^2)&\x&\U(2)&\longrightarrow&\SU(2)&\x&\U(1)_+\cr
\hbox{\scriptsize Lorentz}
&&\hbox{\scriptsize hypercharge-isospin}&&\hbox{\scriptsize spin}&&
\hbox{\scriptsize electromagnetic}\cr
\end{array}
\end{eq}For example, the  lepton field
in the minimal model connects, for  each spacetime translation,
the two $\SL(\C^2)$-degrees of freedom 
with the two isospin $\SU(2)$ degrees of
freedom and  a hypercharge $\U(1)$ value $y=-{1\over2}$
\begin{eq}{l}
\R^4\ni x\mape\bl l(x)_\al^A\hbox{ with }A,\al\in\{1,2\}.
\end{eq}The transition from interaction field to particles 
with respect to internal degrees of freedom uses the ground state degeneracy,
implemented by the $\U(2)$-invariant condition 
 $\lrangle{\Phi^\star \Phi(x)}=M^2>0$
with  the Lorentz scalar Higgs  field
\begin{eq}{l}
\R^4\ni x\mape\Phi(x)^\al.
\end{eq}It is an isospin doublet with hypercharge $y={1\over 2}$.
The Higgs field  transmutes from ``general" hyperisospin $\U(2)$-properties 
 of the lepton field to  ``idolized" electromagnetic 
$\U(1)_+$-properties of the particles,
e.g.,  for the electron-positron field, an isosinglet
  with electromagnetic charge number $z=-1$, 
\begin{eq}{l}
\U(2)\map\U(1)_+:~~\bl l(x)_\al^A\mape \bl e(x)^A= 
{\Phi(x)^\al\over 
|\Phi|(x)}
\bl l(x)_\al^A=
 \bl l(x)_2^A+\dots 
\end{eq}The ``idolization" 
comes with the distinction of a ground state and  
the expansion of the Higgs transmutator (more below) ${\Phi^{\al}(x)\over
|\Phi|(x)}=\de^\al_2+\dots$ for
$e^2={\scriptsize\begin{pmatrix}0\cr1\cr\end{pmatrix}}\cong\de_2^\al$ 
and $|\Phi|(x)=\sqrt{\Phi^\star \Phi(x)}$.

With respect to external  degrees of freedom, the 
transition from  a left-handed 
Weyl  field with Lorentz group $\SL(\C^2)$-action to particles 
with mass $m>0$ and $\SU(2)$-spin 
requires a rest system.
The related harmonic expansion of the spacetime field 
with respect to eigenvectors involves the
electron creation and positron annihilation  operators 
$\ro u^a(\rvec q)$ and $\ro a^{\star a}(\rvec q)$
respectively for spin directions 
$a\in\{1,2\}$
and  mo\-men\-tum $\rvec q$ as translation  eigenvalues, 
\begin{eq}{rl}
\SL(\C^2)\map \SU(2):&\bl e(x)^A\mape \ro u(\rvec q)^a,~\ro a^{\star}(\rvec q)^a \cr
\hbox{where}&\bl e(x)^A=\plint{d^3 q\over 2 q_0}s({q\over m})^A_a~
[e^{iqx}\ro u(\rvec q)^a+e^{-iqx}\ro a^{\star}(\rvec q)^a]\cr
&\hbox{ with } q_0=\sqrt{m^2+\rvec q^2}.\cr
\cr
\end{eq}The boost re\-pre\-sen\-ta\-tion $s({q\over m})^A_a$, 
discussed below as Weyl transmutator, connects
the Lorentz group $\SL(\C^2)$-action  for fields with a rest system
spin $\SU(2)$-action for massive particles.

Altogether, there are four transmutators involved with 
$G\x H$-trans\-for\-ma\-tions for 
four different group pairs $H\subnoteq G$: the lepton field
with external-internal trans\-for\-ma\-tion behavior, the Higgs field as internal
transmutator from interaction to particles,
the boost representation as corresponding external transmutator, which leaves
the creation and annihilation operators with the external-internal properties of
the particles (spin and charge)
\begin{eq}{c}

\begin{array}{rrcll}
&&\hbox{interactions}&&\cr
&&{\scriptstyle \bl l(x)_\al^A}&&\cr
&\GL(\C^2)&\longleftrightarrow&\U(2)&\cr
\noalign{\vskip 1mm}
\hbox{external}&{\scriptstyle s({q\over m})^a_A}\updownarrow\hskip4mm
&&\hskip3mm\updownarrow{\scriptstyle \Phi(x)^\al}&\hbox{internal}\cr
\noalign{\vskip 2mm}
&\SU(2)&\longleftrightarrow&\U(1)_+&\cr
&&{\scriptstyle \ro u(\rvec q)^a, ~\ro a^\star(\rvec q)^a}&&\cr
&&\hbox{particles}&&\cr

\end{array}

\end{eq}

\subsection{Pauli  transmutator}

Perpendicular relativity, pa\-ra\-me\-tri\-zable by a  2-sphere of radius $r$,
is re\-pre\-sen\-ted 
as linear operator by the 
fundamental Pauli transmutator from   rotations to axial  rotations, 
\begin{eq}{rl}
\R^3\supnoteq\Om^2\ni {\rvec x\over r}&\mape u({\rvec x\over r})\in\SU(2),\cr 
u({\rvec x\over r})\o\si_3\o u^\star({\rvec x\over r})&=
{\rvec  x\over r}={\si_ax_a\over r}\hbox{ with }r^2=\rvec x^2,\cr
u({\rvec x\over r})= e^{i\rvec\al}&=\bl1_2\cos \al+i
{\rvec\al\over \al}\sin \al\hbox{ with }
\tan2\al=\tan\th={\sqrt{x_1^2+x_2^2}\over x_3}
\cr
&=
\sqrt{{x_3+r\over 2r}}
\brack{\bl1_2+i{\rvec x_\perp\over x_3+r}}
= {{1\over\sqrt{2r(x_3+r)}}}
{\scriptsize
\left(\begin{array}{cc}
x_3+r&-x_1+ix_2\cr
 x_1+ix_2& x_3+r\cr\end{array}\right)}\cr
=u(\phi,\th)&={\scriptsize\begin{pmatrix}
\cos{\th\over2}&-e^{-i\phi}\sin{\th\over2}\cr
e^{i\phi}\sin{\th\over2}&\cos{\th\over2}\cr\end{pmatrix}}.
 
\cr
\end{eq}

An action on the Pauli transmutator   $u({\rvec x\over r })$
 from left with the spin group 
$\SU(2)$ 
gives the transmutator 
at the rotated point $O.\rvec x$ on the 2-sphere and a right action 
with  the axial group $\SO(2)$ (Wigner axial rotation)
\begin{eq}{rl}
o\in \SU(2):& 
o\o u({\rvec x\over r})=
u({O.\rvec x\over r })\o v(o,{\rvec x\over r })\cr
\hbox{with }&\left\{\begin{array}{l}
v(o,{\rvec x\over r }) \in \SO(2),\cr
O.\rvec x= o\o \rvec x\o o^\star,\cr
O_a^b={1\over 2}\tr \si_a\o o\o \si_b\o o^\star\in\SO(3).\cr
\end{array}\right.

\end{eq}The explicit complicated looking expression for the  Wigner
axial rotation  can be computed from 
$v(o,{\rvec x\over r })
 =u^\star({O.\rvec x\over r })\o o\o u({\rvec x\over r})$. 

\subsection{Weyl transmutators}

In special relativity, the Weyl re\-pre\-sen\-ta\-tions of the boosts, pa\-ra\-me\-tri\-zed by 
the en\-er\-gy-mo\-men\-tum hyperboloid for mass $m>0$,
are a familiar example for a transmutator,
\begin{eq}{rl}
\R^4\supnoteq\cl Y^3\ni{q\over m}&\mape   s({q\over m})\in\SL(\C^2),\cr
s({q\over m})\o\bl1_2\o  s^\star( {q\over m})&={ q\over m}={\si^jq_j\over m}\hbox{ with
}m^2=q^2\cr
\end{eq}and Weyl matrices $\si^j=(\bl 1_2,\rvec \si)$
and  $\d\si^j=(\bl 1_2,-\rvec \si)$. The explicit expressions
involve the Pauli transmutator for the two spherical degrees of freedom:
\begin{eq}{rl}
{ q\over m}&
=u({\rvec q\over|\rvec q| })
\o e^{2\be\si_3}\o 
u^\star({\rvec q\over |\rvec q| }),\cr
e^{2\be\si_3}&=\diag { q\over m}=
{1\over m}{\scriptsize
\left(\begin{array}{cc}
q_0+|\rvec q|&0\cr
0&q_0-|\rvec q|\cr\end{array}\right)},~~\tanh2\be=
{|\rvec q|\over  q_0 }={v\over c},\cr

s({q\over m})&=u({\rvec q\over|\rvec q| })
\o e^{\be\si_3} =
 \bl 1_2\cosh \be+ {\rvec q\over|\rvec q|}\sinh\be\cr&=\sqrt{{ q_0 +m\over2m}}
\Brack{\bl 1_2+{\rvec q\over  q_0 +m}}
 
 = {{1\over\sqrt{2m(q_0+m)}}}
{\scriptsize
\left(\begin{array}{cc}
q_0+q_3+m&-q_1+iq_2\cr
 q_1+iq_2& q_0-q_3+m\cr\end{array}\right)}.\cr
\end{eq}

The left-handed Weyl transmutator $s({q\over m})\in\SL(\C^2)$ 
together with its right-handed partner 
$\hat s({q\over m})=u({\rvec q\over|\rvec q| })
\o e^{-\be\si_3}\in\SL(\C^2)$
where $\hat s= s^{-1\star}$ are the two fundamental  transmutators
from Lorentz group to rotation subgroups.
The restriction in the en\-er\-gy-mo\-men\-ta  from four to three parameters by 
the on-shell hyperboloid $\cl Y^3$ condition ${q^2\over m^2}=1$ is expressed
by the Dirac equation in en\-er\-gy-mo\-men\-tum space,
\begin{eq}{l}
\left.\begin{array}{rlrl}
s({q\over m})\o \hat s^{-1}( {q\over m})&={\si^jq_j\over m}&\then
s({q\over m})&={\si^jq_j\over m}\o \hat s( {q\over m})\cr

\hat s({q\over m}) \o s^{-1}( {q\over m})&={\d\si^jq_j\over m}&\then
\hat s({q\over m}) &={\d\si^jq_j\over m}\o s( {q\over m})\end{array}
\right\}
\then (\ga^jq_j-m)\bl s({q\over m})=0\cr\cr
\hbox{ with }
\ga^j
={\scriptsize\begin{pmatrix}0&\si^j\cr \d\si^j&0\cr\end{pmatrix}},~~
\bl s({q\over m})
={\scriptsize\begin{pmatrix}s({q\over m})& 0\cr0&\hat s({q\over m})
\cr\end{pmatrix}}.

\end{eq}The four columns of the $(4\x4)$ matrix 
$\bl s({q\over m})$ are familiar as solutions of the Dirac equation.  

For the Pauli transmutator, the  analogue to the 
 Dirac equation is the condition 
$\si^ax_a u({\rvec x\over r})-u({\rvec x\over r})\si_3r=0$,
which restricts the three parameters 
to two independent perpendicular ones.

An action from left with the Lorentz group 
$\SL(\C^2)$ gives the Weyl transmutator 
at the Lorentz transformed en\-er\-gy-mo\-men\-ta $\La.q$ on the 
hyperboloid $q^2=m^2$, accompanied 
by a right action with a Wigner spin $\SU(2)$-rotation 
\begin{eq}{rl}
\la\in  \SL(\C^2):&\la\o s({q\over m})=s({\La.q\over m})\o
u({q\over m},\la)\cr
\hbox{with }&
\left\{\begin{array}{l}
u({q\over m},\la)\in\SU(2),\cr
\La. q=\la\o q\o \la^\star,\cr
\La_j^k={1\over 2}\tr\si_j\o \la\o 
\d\si^k\o  \la^\star\in\SO_0(1,3).
\end{array}\right.
\end{eq}


\subsection{Higgs transmutators}

In the standard model of electroweak interactions
 the three weak  parameters
for the Goldstone manifold of electromagnetic relativity are implemented 
by three chargelike degrees of freedom of   the Higgs vector
$\Phi^\al\cong{\scriptsize\begin{pmatrix}\Phi^1\cr\Phi^2\cr\end{pmatrix}}\in V\cong\C^2$  
and its orthogonal $\ep^{\al\be}\Phi^\star_\be$,
\begin{eq}{rl}
\C^2\supnoteq\cl G^3\ni {\Phi\over M}&\mape v({\Phi\over M})\in\U(2),\cr 
v({\Phi\over M})&=
{\scriptsize\begin{pmatrix}
e^{i(\al_3-\al_0)}\cos{\th\over2}&- e^{-i(\phi-\al_0)}\sin{\th\over2}\cr
e^{i(\phi-\al_0)}\sin{\th\over2}&e^{-i(\al_3-\al_0)}\cos{\th\over2}\cr\end{pmatrix}}
=u(\phi-\al_3,\th)\o e^{i(\al_3-\al_0)\si_3 }
\cr
&={1\over M}{\scriptsize\left(\begin{array}{cc}
\Phi^\star_ 2&\Phi^1\cr
-\Phi^\star_ 1&\Phi^2\cr\end{array}\right)}
\hbox{ with }\det v({\Phi\over M})={|\Phi^1|^2+|\Phi^2|^2\over M^2}=1.\cr

\end{eq}The restriction 
from four to  three real weak degrees of freedom
uses the $\U(2)$-invariant scalar product
$\sprod\Phi\Phi=M^2$ of the Higgs vector space. 
  
A left hypercharge-isospin action
on the 
fundamental Higgs transmutator
gives the transmutator at the  $\U(2)$-transformed
Higgs vector on the Goldstone manifold, accompanied by
a Wigner  electromagnetic $\U(1)_+$-trans\-for\-ma\-tion
from right
\begin{eq}{rl}
u\in  \U(2):&u\o v({\Phi\over M}) =v({u.\Phi\over M})\o
u_+
\hbox{ with }
\left\{\begin{array}{l}
u=e^{i\ga_0}u_2\in\U(1)\o\SU(2),\cr
u_+={\scriptsize\begin{pmatrix}e^{i2\ga_0}&0\cr 0&1\cr\end{pmatrix}}\in\U(1)_+.\cr
\end{array}\right.
\end{eq}

\subsection{Real tetrads (vierbeins)}

For general relativity, the 10-pa\-ra\-me\-tri\-c  
$\GL(\R^4)$-orbit of the orthonormal
$\O(1,3)$-`idolized " Lorentz metric  in a symmetric matrix 
$\eta=\eta^T$ is diagonalizable
to four principal axes
with a trans\-for\-ma\-tion from a maximal compact subgroup
$\O(4)\subnoteq \GL(\R^4)$ (6 parameters),
\begin{eq}{rl}
 \ga=h\o \eta\o  h^T= \ga^T&=
 O(\ga)\o \diag \ga\o  O(\ga)^T \hbox{ with }O(\ga)\in\O(4).\cr
\end{eq}The diagonal
part of the metric hyperboloid, multiplied by 
the inverse metric $\eta^{-1}$, displays the  
remaining four dilation trans\-for\-ma\-tions 
from the maximal noncompact abelian subgroup, 
\begin{eq}{l}
\eta^{-1}\o \diag \ga=e^{2[d(\ga)+d_0(\ga)]}\in\D(1)\x\SO_0(1,1)^3\cong \D(1)^4\subnoteq
\GL(\R^4).
\end{eq}The diagonal elements are four directional units, one for  time 
and three for the metric ellipsoid of 3-position.

The operational decomposition of
the metric hyperboloid  leads to the pa\-ra\-me\-tri\-zation 
of the 10-dimensional tetrad $h$ as basis of 
real 4-di\-men\-sio\-nal tangent spacetime $\R^4$
with four dilations  and a 6-dimensional rotation 
\begin{eq}{l}
\cl M^{10}\ni \ga\mape h(\ga) \in \D(1)^4\x \O(4)\subnoteq \GL(\R^4)
,~~h(\ga)=e^{d(\ga)+d_0(\ga)}\o O(\ga).
\end{eq}

A general linear $\GL(\R^4)$ left-multiplication gives 
the tetrad for a transformed metric tensor and a Wigner 
right-trans\-for\-ma\-tion by the idolized Lorentz group $\O(1,3)$,  
\begin{eq}{rl}
g\in  \GL(\R^4):&g\o h(\ga) =h(g\o \ga\o g^T)\o \La(g,\ga)
\hbox{ with } \La(g,\ga)\in\O(1,3).
\end{eq}

\subsection{Complex dyads (zweibeins)}

Nonlinear spacetime $\cl D^4$, i.e., 
the orientation manifold of $\U(2)$-scalar products
for unitary relativity,
is pa\-ra\-me\-tri\-zable by the future cone,  
\begin{eq}{l}
 x\in\cl D^4\cong \GL(\C^2)/\U(2)=\D(\bl1_2)\x\SL(\C^2)/\SU(2)\cong \D(1)\x\cl
 Y^3.\cr
\end{eq}It is transformed to an ``idolized" diagonal
scalar product by   a 
Weyl transmutator $s({ x\over \sqrt{x^2}})
\in\SL(\C^2)$  for the three hyperbolic degrees of freedom and 
a dilation $\D(1)=\exp\R\cong\R$ for eigentime 
$e^{2\be_0}=\sqrt {x^2}$, 
\begin{eq}{rl}
 x&={\scriptsize\begin{pmatrix}
x_0+x_3&x_1-ix_2\cr x_1+ix_2&x_0-x_3\cr\end{pmatrix}}
=s({ x\over \sqrt{x^2}})\o e^{2\be_0\bl 1_2}\o s^\star({ x\over \sqrt{x^2}})
=u({\rvec x\over r })
\o\diag  x\o 
u^\star({\rvec x\over r }),\cr
\diag  x&={\scriptsize\begin{pmatrix}
x_0+  r &0\cr 0&x_0- r \cr\end{pmatrix}}
=e^{2(\be_0\bl 1_2+\be\si_3)}\in\D(1)\x\SO_0(1,1)\cr
&\hbox{with }e^{4\be_0}={x^2},~\tanh 2\be={r\over x_0}.\cr
\end{eq}

The diagonalization of the scalar products gives the fundamental transmutator
from the extended Lorentz group
to the hyperisospin subgroup. It is a basis of the complex 2-di\-men\-sio\-nal space
and will be called, in analogy to a real tetrad or vierbein, 
a complex  dyad or zweibein. It is  pa\-ra\-me\-tri\-zed
by the  future cone spacetime points as orbit of the $\U(2)$-scalar product,  
\begin{eq}{rl}
\R^4\supnoteq \cl D^4\ni  x&\mape \psi( x)\in\GL(\C^2), \cr
\psi( x)\o\bl1_2\o \psi^\star( x)&= x,\cr

\psi(x)&=s({ x\over \sqrt{x^2}})\o e^{\be_0\bl 1_2}
=u({\rvec x\over r })\o e^{\be_0\bl 1_2+\be\si_3}.\cr

\end{eq}

The left  action with the extended Lorentz group $\GL(\C^2)$ 
as external trans\-for\-ma\-tion gives the dyad $\psi$ 
at a Lorentz transformed and dilated spacetime point in the future cone,
accompanied by an action
 from right with an internal spacetime dependent Wigner
 hyperisospin 
$\U(2)$-trans\-for\-ma\-tion, 
\begin{eq}{rl}
g\in  \GL(\C^2):&g\o \psi( x)=\psi(e^{2\de_0}\La. x)\o
u( x,g)\cr\hbox{with }&
\left\{\begin{array}{l}
u( x,g)\in\U(2),\cr
g=e^{\de_0+i\al_0}\la\in\D(1)\x\U(1)\o\SL(\C^2), \cr
g\o x\o g^\star=e^{2\de_0}\la\o x\o \la^\star
=e^{2\de_0}\La. x,~~\La\in\SO_0(1,3). 
\end{array}\right.
\end{eq}

\section{Linear  re\-pre\-sen\-ta\-tions of relativities}

In the foregoing section, the classes $G/H$  of the five relativities
with  linear groups 
were represented by defining linear trans\-for\-ma\-tions.
The products of these fundamental transmutators
give the finite-di\-men\-sio\-nal re\-pre\-sen\-ta\-tions
of the homogeneous spaces $G/H$.

\subsection{Rectangular transmutators}

Re\-pre\-sen\-ta\-tions of the ``general" group $G$
involve re\-pre\-sen\-ta\-tions of the
cosets $G/H$ re\-pre\-sen\-tatives, 
\begin{eq}{rl}
G\ni g&\mape D(g)\in \GL(V),\cr
G/H\ni gH\ni  g_r&\mape D(g_r),
\end{eq}e.g. for perpendicular relativity
\begin{eq}{rl}
\SU(2)\ni 
u(\phi,\th,\chi)
&={\scriptsize\begin{pmatrix}
e^{i{\chi+\phi\over2}}\cos{\th\over2}&
-e^{i{\chi-\phi\over2}}\sin{\th\over2}\cr
e^{-i{\chi-\phi\over2}}\sin{\th\over2}&
e^{-i{\chi+\phi\over2}}\cos{\th\over2}\end{pmatrix}}
\cr\mape u(\phi,\th,\chi)_r= u(\phi,\th)&={\scriptsize\begin{pmatrix}
\cos{\th\over2}&
-e^{-i\phi}\sin{\th\over2}\cr
e^{i\phi}\sin{\th\over2}&
\cos{\th\over2}\end{pmatrix}}\rin\SU(2)/\SO(2)\cong\Om^2
\end{eq}

A $G$-re\-pre\-sen\-ta\-tion
can be  decomposed into $H$-re\-pre\-sen\-ta\-tions.
In an $(n\x n)$ matrix re\-pre\-sen\-ta\-tion,
\begin{eq}{l}
D(g)\in V\ox V^T\cong\C^n\ox\C^n
\stackrel{\rm e.g., }\sim {\scriptsize\begin{pmatrix}
\m&\m&\m&\m&\m&\m&\m&\m\cr
\m&\m&\m&\m&\m&\m&\m&\m\cr
\m&\m&\m&\m&\m&\m&\m&\m\cr
\m&\m&\m&\m&\m&\m&\m&\m\cr
\m&\m&\m&\m&\m&\m&\m&\m\cr
\m&\m&\m&\m&\m&\m&\m&\m\cr
\m&\m&\m&\m&\m&\m&\m&\m\cr
\m&\m&\m&\m&\m&\m&\m&\m\cr
\end{pmatrix}}\cr
\end{eq}with an $(8\x 8)$-example for $\SU(3)$
and the octet decomposition $8=2+1+3+2$
into $\SU(2)$-representations,
\begin{eq}{l}
V \stackrel
H\cong{\PL_{\io=1}^k}V_\io,~~H\m V_\io\sub V_\io,~~
D(h)\stackrel H\cong{\PL_{\io=1}^k}d_\io(h)
\stackrel{\rm e.g., }\sim {\scriptsize\left(\begin{array}{cc|c|ccc|cc}
\m&\m&0&0&0&0&0&0\cr
\m&\m&0&0&0&0&0&0\cr\hline
0&0&\m&0&0&0&0&0\cr\hline
0&0&0&\m&\m&\m&0&0\cr
0&0&0&\m&\m&\m&0&0\cr
0&0&0&\m&\m&\m&0&0\cr\hline
0&0&0&0&0&0&\m&\m\cr
0&0&0&0&0&0&\m&\m\cr
\end{array}\right)},\cr

\end{eq}the $G$-re\-pre\-sen\-ta\-tion matrices can be decomposed into 
 rectangular $(n\x n_\io)$ matrices, $n_\io\le n$,
\begin{eq}{l}
D(g)={\PL_{\io=1}^k}D_\io(g)=\(D_1(g)\|D_2(g)\|\cdots\|D_I(g)\)
\stackrel{\rm e.g., }\sim {\scriptsize\left(\begin{array}{cc|c|ccc|cc}
\m&\m&\m&\m&\m&\m&\m&\m\cr
\m&\m&\m&\m&\m&\m&\m&\m\cr
\m&\m&\m&\m&\m&\m&\m&\m\cr
\m&\m&\m&\m&\m&\m&\m&\m\cr
\m&\m&\m&\m&\m&\m&\m&\m\cr
\m&\m&\m&\m&\m&\m&\m&\m\cr
\m&\m&\m&\m&\m&\m&\m&\m\cr
\m&\m&\m&\m&\m&\m&\m&\m\cr
\end{array}\right)}\cr
\end{eq}with 
left-right $G\x H$-action, e.g., the $\SU(3)\x\SU(2)$-action on
octet-dublet, octet-singlet, octet-triplet and octet-dublet. As mappings of the coset re\-pre\-sen\-tatives
$(G/H)_r$,
they  are called transmutators: 
\begin{eq}{l}
(G/H)_r\ni  g_r\mape ~
D_\io(g_r)\in V\ox
V_\io^T\cong\C^n\ox\C^{n_\io}
\cr\hbox{ with }
\left\{\begin{array}{rl}		
D_\io(g_rh)&=D_\io(g_r)\o d_\io(h),~h\in H,\cr
D_\io(kg_r)&=D(k)\o D_\io(g_r)\cr
&=  D((kg)_r)\o d_\io(h(k,g_r)),~k\in G.\cr
\end{array}\right.
\end{eq}With bases of the $G$-vector spaces $\rstate{D;j}\in V$ and 
the $H$-vector spaces $\rstate{\io;a}\in V_\io$ one has in a  Dirac
notation with kets for vectors $\rstate{~~}\in V$ and bras 
for linear forms $\lstate{~~}\in V_\io^T$
\begin{eq}{rl}
 V\ox V_\io^T&\ni D_\io(g_r)
=\rstate{D;j} D_\io(g_r)^j_a\lstate{\io;a},\cr
\hbox{e.g. }\C^8\ox\C^2&\ni D_2(g_r)
=\rstate{8;j} D_2(g_r)^j_a\lstate{2;a},~j=1.\dots,8;~ a=1,2.\cr
\end{eq}

The finite-di\-men\-sio\-nal transmutators are 
$(n\x n_\io)$-di\-men\-sio\-nal vector spaces 
with  $G\x H$-re\-pre\-sen\-ta\-tions. Those re\-pre\-sen\-tations are 
 unitary, called Hilbert re\-pre\-sen\-ta\-tions,  only for the
compact relativities,  i.e.,  in the examples above,  for perpendicular and electromagnetic relativity.
There, the transmutators are complete for the harmonic analysis
of the Hilbert spaces with the square integrable functions $L^2(G/H)$
of the orientation  manifold
 of the relativity (more below).

\subsection{Re\-pre\-sen\-ta\-tions of perpendicular relativity}

For perpendicular relativity, all transmutators from rotations to axial rotations
 arise by the totally symmetric products, denoted by {\scriptsize${\OD^{2J}}$},
   of the fundamental 
  Pauli transmutator $u({\rvec x\over r })\in\SU(2)$,
\begin{eq}{rl}
 \SU(2)/\SO(2)\cong\Om^2&\map\SU(1+2J,)\cr
 {\rvec x\over r }&\mape
[2J]({\rvec x\over r })=\hbox{\scriptsize $\OD^{2J}$}u({\rvec x\over r }),~
\rvec x^2=r^2.\cr
\end{eq}The irreducible spin $\SU(2)$-re\-pre\-sen\-ta\-tions 
$[2J]$ are decomposable into axial rotation 
$\SO(2)$-re\-pre\-sen\-ta\-tions $(n)$  
with dimension 2  for $n\ne0$ and two polarizations $\pm n$
(left- and right-circulary polarized) :
\begin{eq}{l}
\irrep\SU(2)\ni [2J]\stackrel{\SO(2)}\cong
\left\{\begin{array}{rl}
{\PL_{n=0,2,..}^{2J}}(n)&\hbox{for }J=0,1,\dots,\cr
{\PL_{n=1,3,..}^{2J}}(n)&\hbox{for }J={1\over2},{3\over 2},\dots,\cr
\end{array}\right.
\end{eq}e.g.,  for rotations acting on 3-position $\R^3$
with $a,b\in\{1,2,3\}$ and $\al,\be\in\{1,2\}$:
\begin{eq}{rl}
[2]({\rvec x\over r })\cong O({\rvec x\over r })^b_a&={1\over2}
\tr u({\rvec x\over r })\o \si^b\o  u^\star({\rvec x\over r })\o \si^a
\cr&={1\over r }
{\scriptsize \left(  
\begin{array}{c|c}
\de^{\al\be} r -{x_\al x_\be\over r +x_3}&x_\al\cr
-x_\be&x_3\cr\end{array}\right)}\in\SO(3),\cr
[2]&\stackrel{\SO(2)}\cong(2)\pl(0)
\end{eq}with the relations for the 
$\SO(3)$
and  $\SO(2)$
metric tensors
\begin{eq}{l}
O({\rvec x\over r })^a_{\al,3}\de_{ab}
O({\rvec x\over r })^b_{\be,3}
={\scriptsize\left(\begin{array}{c|c}
\de_{\al\be}&0\cr\hline
0&1\cr\end{array}\right)},~~
O({\rvec x\over r })^a_\al\de^{\al\be}O({\rvec x\over r })^b_\be
=\de^{ab}-{x_ax_b\over r^2}.\cr
\end{eq}

The 2nd symmetric 
power of the Pauli transmutator, in a Cartesian and a spherical basis,
\begin{eq}{l}
O({\rvec x\over r })
={\scriptsize \left(  
\begin{array}{c|c}
\de^{\al\be} -{x_\al x_\be\over r( r +x_3)}&{x_\al\over r}\cr
-{x_\be\over r}&{x_3\over r}\cr\end{array}\right)}
\cong
{\scriptsize \left(  
\begin{array}{cc|c}
e^{i\phi}\cos^2{\th\over2}&
-e^{i\phi}\sin^2{\th\over2}&
ie^{i\phi}{\sin\th\over\sqrt2}\cr
i{\sin\th\over\sqrt2}&
i{\sin\th\over\sqrt2}&\cos\th\cr
-e^{-i\phi}\sin^2{\th\over2}&
e^{-i\phi}\cos^2{\th\over2}&
ie^{-i\phi}{\sin\th\over\sqrt2}\cr

\end{array}\right)},
\end{eq}displays in the 3rd column $O({\rvec x\over r })_3^b$ 
 the spherical harmonics $\ro Y_1(\phi,\th)\sim{\rvec x\over
r}$ as a basis for the $\C^3$-Hilbert subspace in the  Hilbert space 
$L^2(\Om^2)$ with the square integrable functions on the 2-sphere.
Its symmetric traceless products of power $J=1,2,\dots$ give 
the spherical harmonics $\ro Y_J(\phi,\th)\sim({\rvec x\over
r})^J_{\rm traceless}$ which arise as the $(1+2J)$-entries in one column of 
the $(1+2J)\x(1+2J)$ matrices for the re\-pre\-sen\-tation $[2J]$. 
The spherical harmonics are  bases for the Hilbert spaces $\C^{1+2J}\subnoteq L^2(\Om^2)$
with the irreducible $\SO(3)$-re\-pre\-sen\-ta\-tions.

With respect to the dichotomic $\SU(2)\x\SO(2)$-trans\-for\-ma\-tion behavior, 
the four functions in the two columns of the $(2\x2)$-Pauli transmutator,
\begin{eq}{l}
u({\rvec x\over r})
=
{\scriptsize
\left(\begin{array}{cc}
\sqrt{{x_3+r\over 2r}}&-{x_1-ix_2\over\sqrt{2r(x_3+r)}}\cr
 {x_1+ix_2\over\sqrt{2r(x_3+r)}}&\sqrt{{x_3+r\over 2r}}\cr\end{array}\right)}
={\scriptsize
\left(\begin{array}{cc}
\cos{\th\over2}&-e^{-i\phi}\sin{\th\over2}\cr
e^{i\phi}\sin{\th\over2}&\cos{\th\over2}\cr\end{array}\right)}\in\C^2\ox\C^2, 
\end{eq}and the six functions in the 1st and 2nd column  
$O({\rvec x\over r})^b_{1,2}\cong O({\rvec x\over r})^b_{+,-}$ above
in a rectangular $(3\x 2)$ matrix
constitute bases for  finite-di\-men\-sio\-nal Hilbert spaces
$\C^2\ox\C^2$ and $\C^3\ox\C^2$
with $\SU(2)$-re\-pre\-sen\-ta\-tions on $\C^2$ and $\C^3$,
acting from left, and nontrivial $\SO(2)$-re\-pre\-sen\-ta\-tions
$\SO(2)\ni e^{i\al_3\si_3}\mape e^{in\al_3\si_3}$ on $\C^2$
with $n=1,2$ respectively,
 acting from right. They are  irreducible subspaces in the
harmonic analysis 
of the  Hilbert space $L^2(\Om^2,\C^2)$ with the square 
integrable mappings from the 2-sphere into a vector space 
with  nontrivial $\SO(2)$-action.

In general one has the Peter-Weyl decompositions \cite{PWEYL} 
into irreducible subspaces for $\SU(2)\x\SO(2)$ action: 
\begin{eq}{l}
V_{n}\cong\C^{2-\de_{n0}}:~~
L^2(\Om^2,V_{n})\cong{\PL_{2J\ge n}\C^{1+2J}}\ox V_{n}\hbox{ (dense).}
\end{eq}The orthogonal  sum goes over all 
$\SU(2)$-re\-pre\-sen\-ta\-tion that contain the 
$\SO(2)$-re\-pre\-sen\-ta\-tion
on $V_{n}\cong\C,\C^2$.
This generalizes the case for the spherical harmonics with $V_0\cong\C$.

\subsection{Re\-pre\-sen\-ta\-tions of rotation relativity} 

For special relativity, all finite-di\-men\-sio\-nal 3-hy\-per\-bo\-loid re\-pre\-sen\-ta\-tions 
(boost re\-pre\-sen\-ta\-tions), i.e.,  all 
finite-di\-men\-sio\-nal transmutators from 
Lo\-rentz group to  ro\-ta\-tion group, 
 can be built by the totally symmetric products of the two fundamental 
Weyl transmutators $s({q\over m}),\hat s({q\over
m})\in\SL(\C^2)$,
 \begin{eq}{rl}
\SL(\C^2 )/\SU(2)\cong\cl Y^3&\map\SL(\C^{(1+2L)(1+2R)}),\cr
{q\over m}&\mape
[2L|2R]({q\over m}) =
\hbox{\scriptsize $\OD^{2L}$}
s({q\over m})\ox\hbox{\scriptsize $\OD^{2R}$}\hat s({q\over m}),~~q^2=m^2.\cr
\end{eq}The finite-di\-men\-sio\-nal 
irreducible Lorentz group re\-pre\-sen\-tations can be
decomposed into irreducible
spin  re\-pre\-sen\-tations,
\begin{eq}{l}
\irrep^{\rm finite}\SL(\C^2)\ni
[2L|2R]\stackrel{\SU(2)}\cong{\PL_{J=|L-R|}^{L+R}}[2J].
\end{eq}
 
For example, the vector re\-pre\-sen\-ta\-tion $\La=[1|1]$
gives two irreducible transmutators  from Lo\-rentz group
 to  ro\-ta\-tion group,
the first column for spin 0-representation and the three remaining columns
for spin 1-representation, with $a,b\in\{1,2,3\}$,
\begin{eq}{rl}
[1|1]({q\over m})=\La({q\over m})^j_k&\cong{1\over2}
\tr s({q\over m})\o \si^j \o s^\star({q\over m})\o \d\si_k
\cr&={{1\over m}}
{\scriptsize\left(\begin{array}{c|c}
q_0&q_a\cr
q_b&\de_{ab} m+{q_aq_b\over m+q_0}\cr\end{array}\right)}\in\SO_0(1,3),\cr
[1|1]&\stackrel{\SO(3)}\cong[0]\pl[2].\end{eq}The four columns of the matrix
$\La({{q\over m}})^j_{0,a}$
relate to each other the metric tensors of $\SO_0(1,3)$ and $\SO(3)$
\begin{eq}{l}
\La({ {q\over m}})^k_{0,a}\eta_{kj}\La({ {q\over m}})^j_{0,b}=
{\scriptsize\left(\begin{array}{c|c}
1&0\cr\hline
0&-\de_{ab}\cr\end{array}\right)},~~
\La({ {q\over m}})^k_a\de^{ab}
\La({ {q\over m}})_b^j=-\eta^{kj}+{q^kq^j\over m^2}.
\end{eq}

The 
transmutators from 
Lo\-rentz group to  ro\-ta\-tion group 
in the rectangular $(4\x 3)$-submatrix
$\La({ {q\over m}})^k_a\in\R^4\ox \R^3$  are used for massive spin 1 particles in  Lorentz 
vector fields, e.g.,  
for the neutral weak boson
and its Feynman propagator,
 \begin{eq}{l}
\bl Z(x)^j
=\plint {d^3q\over 2q_0} 
\La({q\over m})^j_a
[  e^{iqx}\ro u(\rvec q)^a+ e^{-iqx}\ro u^\star(\rvec q)^a],
\cr
\lrangle{\acom{\bl Z^k(y)}{\bl Z^j(x)}
-\ep(x_0-y_0)\com{\bl Z^k(y)}{\bl Z^j(x)}}
=  {i\over\pi}\int{d^4q\over(2\pi)^3}
{(-\eta^{kj}+{q^kq^j\over m^2})\over q^2+io-m^2} e^{iq(x-y)}.\cr
\end{eq}

In contrast to compact perpendicular relativity
with the Hilbert space $L^2(\Om^2)$,
the Hilbert space for the square integrable functions on the
 3-hy\-per\-bo\-loid $L^2(\cl Y^3)$
has no finite-di\-men\-sio\-nal Hilbert subspaces with irreducible 
 $\SL(\C^2)$-re\-pre\-sen\-ta\-tions. The monomials in the
columns of the fundamental Weyl transmutators give bases for
finite-di\-men\-sio\-nal $\SL(\C^2)\x\SU(2)$ 
re\-pre\-sen\-ta\-tions on $\C^{(1+2L)(1+2R)}\ox\C^{1+2J}$,
which are indefinite unitary 
for the noncompact Lorentz  group $\SL(\C^2)$.
The spin $\SU(2)$-re\-pre\-sen\-ta\-tion has to be 
contained in 
 the $\SL(\C^2)$-re\-pre\-sen\-ta\-tion, i.e.,  $|L-R|\le J\le L+R$.

\subsection{Re\-pre\-sen\-ta\-tions of electromagnetic relativity} 

For the standard model of electroweak interactions, 
the Higgs pa\-ra\-me\-tri\-zed defining  re\-pre\-sen\-ta\-tion 
of the orientation manifold  $\cl G^3$ of the electromagnetic group
with hypercharge $y={1\over2}$ and  isospin $T={1\over2}$
and its conjugate, i.e.,  the two Higgs transmutators
\begin{eq}{rl}
[{1\over 2}|1]({\Phi\over  M })&= v({\Phi\over  M })
={1\over M}{\scriptsize\left(\begin{array}{c|c}
\Phi^\star_ 2&\Phi^1\cr
-\Phi^\star_ 1&\Phi^2\cr\end{array}\right)},~~|\Phi|^2=M^2,\cr

[-{1\over 2}|1]({\Phi\over  M })&=v^\star({\Phi\over  M }),
\end{eq}give, via their  products, 
all irreducible re\-pre\-sen\-ta\-tions
of the Goldstone manifold:
\begin{eq}{rl}
\U(2)/\U(1)_+\cong\cl G^3\map\U(1+2T),& 
{\Phi\over  M }\mape [\pm n+T|2T]({\Phi\over  M }),\cr

\irrep\U(2)\ni[\pm n+T|2T]&\cong [\pm 1|0]^n\ox
\hbox{\scriptsize $\OD^{2T}$}[{1\over 2}|1],\cr
\irrep\U(1)\ni[\pm 1|0]&\cong[\pm{1\over 2}|1]\and [\pm{1\over 2}|1].\cr 
\end{eq}Because of the central correlation
$\SU(2)\cap\U(\bl1_2)=\{\pm\bl1_2\}$ in $\U(2)$, 
the $\U(2)$-representations have  
the  correlation of the hypercharge- and isospin-invariant  $y=T\pm n$ with 
natural  $n$,
i.e.,  the two invariants $(y,T)$ 
for the rank 2 $\U(2)$-trans\-for\-ma\-tions are either both integer or both halfinteger
as visible in the colorless fields of the standard model. 

The decomposition of a 
hyperisospin $\U(2)$-re\-pre\-sen\-ta\-tion into  
irreducible re\-pre\-sen\-ta\-tions of the electromagnetic group
 $\U(1)_+\ni e^{i2 \ga_0}\mape e^{zi 2\ga_0}$
is given with integer charge numbers $z\in\Z$:
 \begin{eq}{rl}
\U(2)&\ni
[\pm n+T|2T]\stackrel{\U(1)_+}\cong{\PL_{z=\pm n}^{\pm n+2T}}[z],\cr
\hbox{e.g., }&\left\{\begin{array}{rl}
[\pm{1\over 2}|1]&\hskip-2mm\cong[0]\pl [\pm1],\cr
[0|2]&\hskip-2mm\cong[-1]\pl [0]\pl [1].\cr\end{array}\right.
\end{eq}

The $(1\x 1)$ examples with antisymmetric power 2 of the fundamental Higgs transmutators
give the
transmutation
from hyperisospin $\U(2)$  
to electromagnetic $\U(1)_+$ 
on $\C$ for  hypercharge  nontrivial isospin $\SU(2)$-singlets
with charge numbers $z=\pm1$, 
\begin{eq}{rl}
[1|0]({\Phi\over  M })
&={\tilde\Phi^\star_\al\Phi^\al\over  M ^2}\in\U(1)\hbox{ with }
[1|0]\hskip3mm\stackrel{\U(1)_+}\cong\hskip2mm[1]
,~~\tilde\Phi^\al=\ep^{\al\be}\Phi^\star_\be,\cr
[-1|0]({\Phi\over  M })
&={\Phi^\star_\al\tilde\Phi^\al\over  M ^2}\in\U(1)
\hbox{ with }[-1|0]\stackrel{\U(1)_+}\cong[-1].\cr
\end{eq}The $(3\x 1)$ product of both fundamental Higgs transmutators 
describes an hypercharge  trivial isospin $\SU(2)$-triplet. The columns
are three transmutators to charge $z\in\{-1,0,1\}$,  
\begin{eq}{rl}
[0|2]({\Phi\over  M })
&={1\over 2}\tr \tau^b \o v({\Phi\over  M }) \o \tau^a  \o v^\star({\Phi\over  M })\cr
&=\left(\begin{array}{c|c|c}
{\Phi^\star\rvec\tau\tilde\Phi
+\tilde\Phi^\star\rvec\tau\Phi\over2 M ^2}&
{\Phi^\star\rvec\tau\tilde\Phi
-\tilde\Phi^\star\rvec\tau\Phi\over2i M ^2}&
{\tilde \Phi^\star\rvec\tau\tilde\Phi
-\Phi^\star\rvec\tau\Phi\over2 M ^2}\end{array}\right)\in\SO(3)
\cr
\hbox{ with }&[0|2]\stackrel{\U(1)_+}\cong[-1]\pl[0]\pl [1].\cr
\end{eq}These three transmutators are used
for the transition from the three isospin gauge fields in the
electroweak standard model to the weak boson particles
\begin{eq}{rl}
\tau^a \bl A_a(x) =\bl A(x)&\mape 
v({\Phi(x)\over M})\o\bl A(x)\o v^\star ({\Phi(x)\over M})
+[\p v({\Phi(x)\over M})]\o v^\star ({\Phi(x)\over M})\cr
=&
(\bl W_-(x),\bl W_0(x),\bl W_+(x))
=(\bl A_-(x),\bl A_0(x),\bl A_+(x))+\dots\cr
&\hfill\hbox{ with }[0|2]({\Phi(x)\over M})^a_i=\de^a_i+\dots
\end{eq}For the definition of particles with the transition from 
Lorentz group to rotation group the neutral field $\bl W_0$ is 
combined,  in the Weinberg rotation,  with the hypercharge gauge field.

Similar to perpendicular relativity,
the Hilbert spaces of the square integrable mappings 
of the compact Goldstone manifold  
$L^2(\cl G^3,V_z)$ into a Hilbert space  $V_z\cong\C$
with electromagnetic action $\U(1)_+\ni e^{i2\ga_0}\mape e^{zi2\ga_0}$
 have Peter-Weyl decompositions into finite-di\-men\-sio\-nal subspaces
$\C^{1+2T}\ox\C$ with irreducible  
re\-pre\-sen\-ta\-tions of $\U(2)\x\U(1)$ where the isospin representations
fulfill $2T\ge |z|$.
The representation spaces are given by 
the columns in the products of the fundamental Higgs transmutator
and its conjugate.

\subsection{Re\-pre\-sen\-ta\-tions of unitary relativity} 

All finite-di\-men\-sio\-nal re\-pre\-sen\-ta\-tions of unitary relativity,
i.e., of nonlinear spacetime $\cl D^4$,
\begin{eq}{rl}
\cl D^4\cong\GL(\C^2)/\U(2)\cong\D(1)\x\SO_0(1,3)/\SO(3)&\map
\GL(\C^{(1+2L)(1+2R)}),\cr
\end{eq}use products of the two conjugated dyads, e.g.,
\begin{eq}{rl}
\psi(x)&=u({\rvec x\over r })\o e^{\be_0\bl 1_2+\be\si_3}\in\GL(\C^2),\cr
 \psi(x)\o\bl1_2\o \psi^\star(x)&=
u({\rvec x\over r })\o e^{2(\be_0\bl1_2+\be\si_3)}\o u^\star({\rvec x\over r })
=x\in\GL(\C^2),\cr
(\psi(x)\o\bl1_2\o \psi^\star(x))^2&=e^{4\be_0}={x^2}\in\D(1).\cr
\end{eq}The monomials in 
the dyads span  finite-di\-men\-sio\-nal 
 spaces  with $\GL(\C^2)\x\U(2)$-re\-pre\-sen\-ta\-tions 
that are indefinite unitary 
for the noncompact group
$\GL(\C^2)$. They are  no Hilbert spaces and, therefore, of little importance
for a quantum structure.
Hilbert spaces with faithful representations of nonlinear spacetime have to be
infinite dimensional.
They will be treated in 
``Relativities and homogeneous spaces II
--Spacetime as unitary relativity".

\subsection{Re\-pre\-sen\-ta\-tions of Lorentz group  relativity} 

For general relativity, 
all finite-di\-men\-sio\-nal re\-pre\-sen\-ta\-tions of the general linear group
$\GL(\R^4)$ for the tetrads,
\begin{eq}{l}
\GL(\R^4)/\O(1,3)\cong
\D(1)\x\SL_0(\R^4)/\SO_0(1,3),~~\SL_0(\R^4)\sim\SO_0(3,3),
\end{eq}are obtained by products of 
the fundamental re\-pre\-sen\-ta\-tions
of the rank 3 special subgroup $\SL_0(\R^4)$, which is
locally isomorphic to the indefinite orthogonal group
$\SO_0(3,3)$.
The three fundamental re\-pre\-sen\-ta\-tions
are the two 4-di\-men\-sio\-nal spinor re\-pre\-sen\-ta\-tions, dual to each other, 
and a  6-di\-men\-sio\-nal self-dual one,  
\begin{eq}{l}
\dim_\R[1,0,0]=4,~~
\dim_\R[0,1,0]={4\choose2}=6,~~
\dim_\R[0,0,1]={4\choose3}=4.\cr
\end{eq}The dimensions of the finite-di\-men\-sio\-nal irreducible
$\SL_0(\R^4)$-re\-pre\-sen\-ta\-tions
are given by the Weyl formula: 
\begin{eq}{l}
\irrep^{\rm finite}\SL_0(\R^4)\ni [n_1,n_2,n_3]\cong\N^3,\cr
d_n=\dim_\R[n_1,n_2,n_3]=
{\scriptstyle {(n_1+1)(n_2+1) (n_3+1)
(n_1+n_2+2)(n_3+n_2+2)(n_1+n_2+n_3+3)\over 12}},\cr
\hbox{dual reflection: }[n_1,n_2,n_3]\lrmap[n_3,n_2,n_1].
\end{eq}Self-dual re\-pre\-sen\-ta\-tion spaces, i.e.,  for $[n,m,n]$,  have
an $\SL(\R^4)$-invariant symmetric bilinear form with neutral signature.   

The finite-di\-men\-sio\-nal re\-pre\-sen\-ta\-tions of Lorentz group
relativity,
pa\-ra\-me\-tri\-zed by the metric manifold,
\begin{eq}{rl}
\GL(\R^4)/\O(1,3)\cong\cl M^{10}\ni \ga\mape[n_1,n_2,n_3](\ga)\in\GL(\R^{d_n}),\cr 
\end{eq}have  decompositions with respect to
an ``idolized" Lorentz group.
The three fundamental $\SL_0(\R^4)$-re\-pre\-sen\-ta\-tions give
the two  fundamental $\SO_0(1,3)$-re\-pre\-sen\-ta\-tions,
i.e., the 4-di\-men\-sio\-nal Minkowski re\-pre\-sen\-ta\-tion and the 
6-di\-men\-sio\-nal adjoint re\-pre\-sen\-ta\-tion,
\begin{eq}{rl}
[1,0,0],[0,0,1]&\stackrel{\SO_0(1,3)}\cong[1|1],\cr
[0,1,0]&\stackrel{\SO_0(1,3)}\cong[2|0]\pl[0|2].
\end{eq}

The totally antisymmetric powers of the defining $\SL_0(\R^4)$-re\-pre\-sen\-ta\-tion 
combine the two other fundamental ones
\begin{eq}{l}
{\AND^2}[1,0,0]= [0,1,0],~~
{\AND^3}[1,0,0]= [0,0,1],~~
{\AND^4}[1,0,0]= [0,0,0]\cr
\end{eq}They can be realized by the  tetrad, a spinor re\-pre\-sen\-ta\-tion,
 as fundamental transmutator 
from general linear group to Lorentz group and its totally antisymmetric powers
\begin{eq}{rl}
h_j^\mu(\ga)\in\GL(\R^4),~~h^{lm}_{\ka\la}(\ga)&=
\ep_{\mu\nu\ka\la}\ep^{jklm}h_j^\mu(\ga)h_k^\nu(\ga)\in\GL(\R^6),\cr
h^{m}_{\la}(\ga)&=
\ep_{\mu\nu\ka\la}\ep^{jklm} h_j^\mu(\ga)h_k^\nu(\ga)h_k^\ka(\ga)\in\GL(\R^4)\cr
\det h(\ga)&\in\GL(\R)
\end{eq}The tetrad power-2 product is a  $(6\x 6)$ transmutator acted on by
the self-dual fundamental $\SL_0(\R^4)$-re\-pre\-sen\-ta\-tion
 and the adjoint Lorentz
group re\-pre\-sen\-ta\-tion. It has  
 the same trans\-for\-ma\-tion properties as  the  curvature tensor  
\begin{eq}{l}

\ga\mape R^{lm}_{\ka\la}(\ga)\in \R^6\ox\R^6
~~\hbox{ with }~~
{\scriptsize\begin{array}{r|c}
&[2|0]\pl[0|2]\cr\hline\cr
[0,1,0]&
R^{lm}_{\ka\la}(\ga)\cr\cr\end{array}}
\end{eq}The determinant with power  4 is an $\SL(\R^4)$-scalar 
with nontrivial $\D(1)$-dilation properties.

Obviously, all those real finite-di\-men\-sio\-nal 
re\-pre\-sen\-ta\-tion spaces of the noncompact product $\GL(\R^4)\x\SO_0(1,3)$
have no invariant Hilbert product. A harmonic analysis 
of, e.g., square integrable  functions $L^2(\cl M^{10})$ 
on the metric manifold 
$\cl M^{10}\cong\GL(\R^4)/\SO_0(1,3)$, does
not play a role in classical gravity.


\section{Relativity re\-pre\-sen\-ta\-tions by induction}

Finite-di\-men\-sio\-nal rectangular mappings 
of homogeneous spaces $G/H$
($H$-relativity), as discussed in the foregoing section,
give all Hilbert re\-pre\-sen\-ta\-tion spaces only for compact relativities, 
e.g., 
for perpendicular and electromagnetic relativity.
In general, the  faithful $G\x H$-Hilbert re\-pre\-sen\-ta\-tions 
of a locally compact relativity $G/H$, infinite for noncompact $G$,
can be  induced by re\-pre\-sen\-ta\-tions
of the ``idolized" subgroup $H$.

\subsection{Induced re\-pre\-sen\-tations}

Induced re\-pre\-sen\-tations are 
$G\x H$-subre\-pre\-sen\-tations of the two-sided regular $G\x
G$-re\-pre\-sen\-tation. They are the  extension of the left $G$-action 
$gH\stackrel{_kL}\mape kgH$ on the right $H$-cosets in the form of linear trans\-for\-ma\-tions.

The vector spaces for subgroup $H$-induced $G$-re\-pre\-sen\-tations  
consist of $H$-intertwiners on the group $w:G\map W$ with  values
in a    Hilbert space with a unitary action of the ``idolized" subgroup
$d:H\map\U(W)$.
The $G$-action on the intertwiners is defined by left multiplication $_kL$,
 all this is expressed
in the commutative diagram:
 \begin{eq}{l}
\Diagr{G}{G}WW{_kL\x R_h}{_kw}{d(h)} w,~~\begin{array}{rl}
g,k\in G,~h\in H:&_kL\x R_h(g)=kgh^{-1},\cr
\hbox{$H$-intertwiner:}&w(gh^{-1})=d(h).w(g),\cr
\hbox{$G$-action:}&w\mape {_kw},\cr
&_kw(g)=w(k^{-1}g).\cr
\end{array}
\end{eq}

An $H$-intertwiner on  the group $w\in W^{G/H}$ 
maps  $H$-cosets of the group
into $H$-orbits in the Hilbert space $W$.
It is  defined by its values on re\-pre\-sen\-tatives $g_r\in(G/H)_r\sub G$.
The $G$-action 
comes with the re\-pre\-sen\-ta\-tive dependent $H$-action 
(``gauge group action") of the related Wigner
element  $h(g_r,k)\in H$,
 \begin{eq}{l}
\Diagr{(G/H)_r}{(G/H)_r}WW{_kL}{_kw}{} w,~~\begin{array}{l}
G\x W^{(G/H)_r}\map W^{(G/H)_r},\cr
_{k^{-1}}w(g_r)=w(kg_r)= d(h^{-1}(g_r,k)).w((kg)_r).\cr
\end{array}
\end{eq}The induced re\-pre\-sen\-tation may be
reducible. Since the fixgroups for the left $G$-action on the right
 $H$-cosets are conjugates of  $H$,
\begin{eq}{l}
G_{gH}=\{k\in G\mid kgH=gH\}=gHg^{-1}\cong H,
\end{eq}each  $G$-re\-pre\-sen\-ta\-tion on $W^{(G/H)_r}$ and its subspaces 
contains   the inducing $H$-re\-pre\-sen\-ta\-tion $d$.

With a $G$-left invariant coset measure $dg_r=dkg_r$,
 the intertwiners,  in a bra vector notation,
 have a direct integral expansion with the cosets as natural 
 distributive basis $\lstate {g_r,a}$
and complex coefficients  $w(g_r)_a\in\C$
\begin{eq}{l}
\lstate w
=\plint_{(G/H)_r} dg_r~   w(g_r)_a\lstate {g_r,a}
\in W^{(G/H)_r}.
\end{eq}The $G$-invariant  Hilbert product integrates the Hilbert product of 
the value space $W$
over the cosets
\begin{eq}{l}
W^{(G/H)_r}\x W^{(G/H)_r}\map \C,~~
\norm w^2= \int_{(G/H)_r} dg_r~   \ol{w(g_r)_a}{w(g_r)_a},
\end{eq}An orthonormal distributive basis is defined 
with a Dirac distribution $\de(g_r,g'_r)$, supported by 
the relativity manifold and normalized with respect to the invariant
measure used $dg_r$ (examples below)
\begin{eq}{rl}
\sprod {g'_r,a'}{g_r,a}&=\de_{aa'}\de(g_r,g'_r)\cr
\hbox{ with }
\sprod w{g_r,a}&=
\int_{(G/H)_r}  dg_r'~ \de(g_r,g'_r) w(g'_r)_a=w(g_r)_a.
\end{eq}

In the simplest case, the  functions
on the homogeneous $G$-space for $H$-relativity,  
valued in 
 the complex
numbers as 1-di\-men\-sio\-nal space $W=\C\lstate 1$ 
with trivial $H$-action $d_0(h)=1$, 
are expanded as direct integral over the
cosets  with the corresponding function values
\begin{eq}{l}
\lstate f:(G/H)_r\map\C,~~
 \lstate{f}=   \plint_{(G/H)_r} dg_r~f(g_r)\lstate{g_r}.\cr
\end{eq}They 
are matrix elements (coefficients) of $G$-re\-pre\-sen\-tations
$D$ which 
 contain a trivial $H$-re\-pre\-sen\-tation  
 $D\Sup d_0$.

\subsection{Transmutators as induced re\-pre\-sen\-ta\-tions}

The transmutators above, valued in finite-di\-men\-sio\-nal rectangular matrices,
are acted on with $G\x H$-representations,
a  $G$-action from left, induced by an $H$-action from right
\begin{eq}{rl}
\Diagr{(G/H)_r}{(G/H)_r}{V_D\ox V_\io^T}{V_D\ox V_\io^T}{_kL}{_kD_\io}{} 
{D_\io},&\begin{array}{rl}
_{k^{-1}}D_\io(g_r)&=D_\io(kg_r)=
D(k)\o D_\io(g_r)\cr
&=D((kg)_r)\o d_\io(h(k,g_r)),\cr
\end{array}\cr
\end{eq}

Transmutators can be used for a decomposition
of any $G$-re\-pre\-sen\-ta\-tion
induced by a $H$-re\-pre\-sen\-ta\-tion $d_\io$
on a vector space with basis  $\lstate{\io;a}\in V^T_\io$,
\begin{eq}{l}
w_\io: (G/H)_r\mape V_\io^T,~~
  \lstate{w_\io}=\plint_{(G/H)_r} dg_r~  {w_\io(g_r)_a}\lstate{\io;g_r,a}.\cr

\end{eq}There occur all $G$-re\-pre\-sen\-ta\-tions $D$, which 
contain the inducing $H$-re\-pre\-sen\-ta\-tion $d_\io$. With a basis 
$\rstate {D;j}\in V_D$ one obtains the harmonic $D$-components
$\tilde w_\io(D)_j$, which come with multiplicity $n_D$,
\begin{eq}{l}
\cr \lstate{w^{\rm finite}_\io}={\PL_{D\Sup d_\io}}n_D\tilde w_\io(D)_j~
 \lstate{D^j_\io} 
 \hbox{ with }
 \left\{\begin{array}{rl}
 \lstate{D^j_\io}&= \plint_{(G/H)_r} dg_r~D_\io(g_r)^j_a\lstate{\io;g_r,a},\cr
 \tilde w_\io(D)_j&=\sprod{w_\io}{D;j},\cr
 
 \end{array}\right. 
 \cr
 w_\io^{\rm finite}(g_r)_a={\PL_{D\Sup d_\io}}n_D\tilde w_\io(D)_j 
  D(g_r)_a^j,~~
  w_\io^{\rm finite}(kg_r)_a
  ={\PL_{D\Sup d_\io}}n_D \tilde w_\io(D)_j  D(k)^j_k D(g_r)_a^k,
 
\end{eq}e.g.,  the harmonic analysis of functions  
with the harmonic $D$-components $\tilde f(D)_j$,
\begin{eq}{l}
 \lstate{f^{\rm finite}}={\PL_{D\Sup d_0}}n_D\tilde f(D)_j\lstate{D^j_0}\hbox{ with }
 \left\{\begin{array}{rl}
 \lstate{D^j_0}&= \plint_{(G/H)_r} dg_r~D(g_r)^j_0\lstate{ g_r},\cr
\tilde f(D)_j&=\sprod f{D;j}.\cr
\end{array}\right.  
\end{eq}

\subsection{The Hilbert spaces of compact relativities}

For a compact ``general" group $G$, the
finite-di\-men\-sio\-nal rectangular transmutators 
are square integrable on the manifolds $G/H$. They 
are complete for the harmonic analysis of the group $G$ and its
homogeneous spaces $G/H$, i.e.,  they exhaust
by orthogonal direct Peter-Weyl decompositions with  Schur orthogonality
all square integrable induced 
re\-pre\-sen\-ta\-tions,
\begin{eq}{l}
\hbox{compact }G:~~\left\{\begin{array}{rl}
\lstate{w_\io}&=\lstate{w^{\rm finite}_\io},\cr
L^2(G/H, V_\io^T)&\cong {\PL_{D\Sup d_\io}} n_D~ V_D\ox V_\io^T
\hbox{ (dense),}\cr
\id_{(V_\io^T)^{G/H}}&={\PL_{D\Sup d_\io}} n_D~ \id_{V_D}\cong {\PL_{D\Sup d_\io}} n_D~ \rstate{D;j}\lstate{D;j}.
\end{array}\right.\cr
 \end{eq}There  is Frobenius' reciprocity theorem
for the number $n_D$ of $d_\io$-induced 
$G$-re\-pre\-sen\-ta\-tions $D$.

As discussed above, 
all re\-pre\-sen\-tation matrix  elements of the compact groups
$\U(2)$ and $\SU(2)$
are  square integrable with the finitely decomposable Hilbert spaces
for electromagnetic relativity $L^2(\cl G^3,V_z)$ and  perpendicular
 relativity $L^2(\Om^2,V_{|z|})$. 
 
The complex functions for perpendicular relativity 
are the spherical harmonics\footnote{\scriptsize
In the Euler angle pa\-ra\-me\-tri\-zation, 
both the middle column and the middle row define the
 $\SO(3)$-action on the 2-sphere
$\Om^2\cong\SO(3)/\SO(2)\cong\SO(2)\lq\SO(3)$. The central element  
$\th\mape\cos\th$ pa\-ra\-me\-tri\-zes the double coset space,
the 1-sphere
$\Om^1\cong\SO(2)\lq\SO(3)/\SO(2)\cong\SO(2)$ and is
a spherical $\Om^2$-function.}  
as products of
the three matrix elements 
$\rvec\om\mape {\sqrt{4\pi\over 3}}\ro Y_1(\rvec\om)^a\in\C$
in the middle column 
with trivial  re\-pre\-sen\-tations of $\SO(2)\ni e^{i\chi}$
and a triplet representation of $\SO(3)$,
\begin{eq}{ll}
 {\scriptsize\left(\begin{array}{c|c|c}
e^{i(\chi+\phi)}\cos^2{\th\over2}&
ie^{i\phi}{\sin \th\over\sqrt2}&
-e^{-i(\chi-\phi)}\sin^2{\th\over2}\cr
ie^{i\chi}{\sin\th\over\sqrt2}&
\cos \th&
ie^{-i\chi}{\sin\th\over\sqrt2}\cr
-e^{i(\chi-\phi)}\sin^2{\th\over2}&
ie^{-i\phi}{\sin \th\over\sqrt2}&
e^{-i(\chi+\phi)}\cos^2{\th\over2}\cr\end{array}\right)}\in\SO(3),\cr
\SU(2)/\SO(2)\cong \Om^2\ni{\rvec x\over r}= \rvec \om\mape[2 J](\rvec\om)_0^a=
{\sqrt{4\pi\over 1+2J}}\ro Y_J(\rvec\om)^a\in\C\cr
\hfill\hbox{for } J=0,1,2,\dots\hbox{ with } a\in\{-J,\dots,+J\},\cr
O\in\SO(3): [2 J](O)^a_b\ro Y_J(\rvec\om)^b=\ro Y_J(O.\rvec\om)^a.\cr
\end{eq}There is  Schur's orthogonality  \cite{SCHUR, FOL, KNAPP} 
with the Plancherel normalization given by the
dimension $1+2J$ of the re\-pre\-sen\-ta\-tion space,
\begin{eq}{l}
\int_{\Om^2}{d^2\om\over 4\pi}~\ol{[2J](\rvec\om)^a_0}~[2J'](\rvec\om)^{a'}_0
=\int_{\Om^2}d^2\om~{\ol{\ro Y_J(\rvec\om)^a}\over \sqrt{ 1+2J}}~
{\ro Y_{J'}(\rvec\om)^{a'}\over \sqrt{ 1+2J'}}
={1\over { 1+2J}}\de_{JJ'}\de^{aa'}.\cr
\end{eq}It involves the rotation invariant normalizable
measure and the distributive basis of the 2-sphere,  
\begin{eq}{rl}
\int_{\Om^2}d^2\om&=\int_0^{2\pi}d\phi\int_{-1}^1d\cos\th
=4\pi,\cr
\sprod{\rvec \om'}{\rvec \om}&=\de(\rvec \om-\rvec \om')
=\de(\phi-\phi'){1\over\sin\th}\de(\th-\th').\cr

\end{eq}The spherical harmonics   exhaust  the 
square integrable  2-sphere functions, 
\begin{eq}{rl}
L^2(\Om^2)\ni \lstate f=
\plint_{\Om^2}d^2\om f(\rvec\om)\lstate{\rvec\om}
 &={\PL_{J=0}^\infty}\tilde f(J)_a~
\plint_{\Om^2}d^2\om
 ~\ro Y_J(\rvec\om)^a\lstate{ \rvec\om},\cr
f(\rvec\om)&={\PL_{J=0}^\infty}\tilde f(J)_a~
 ~\ro Y_J(\rvec\om)^a.\cr
\end{eq}The measure  can be rewritten with a 
2-sphere-supported Dirac distribution
\begin{eq}{r}
\lstate{[2J]^a_0}\sim~\plint_{\Om^2}d^2\om
 ~\ro Y_J(\rvec\om)\lstate{ \rvec\om}
=\plint d^3x ~\de(\rvec x^2-1) 
(\rvec x )_{\rm traceless}^{J}\lstate {\rvec x},\cr
\hbox{with }({\rvec x\over |\rvec x|} )_{\rm traceless}^{J}=[2J](\rvec\om).
\end{eq}

\subsection{Finite-di\-men\-sio\-nal analysis of special relativity}

The harmonic analysis of free quantum fields with respect to the
eigenvectors for spacetime translations  and spin rotations
uses  non-Hilbert representations of the Lorentz group. 

Induced re\-pre\-sen\-tations of a  noncompact group
also contain finite-di\-men\-sio\-nal rectangular transmutators. 
For example, the three matrix elements in the middle column with
trivial  re\-pre\-sen\-tations of $\SO(2)\ni e^{i\chi}$,
\begin{eq}{ll}
 {\scriptsize\left(\begin{array}{c|c|c}
e^{i(\chi+\phi)}\cosh^2\be&
e^{i\phi}{\sinh2\be\over\sqrt2}&
-e^{-i(\chi-\phi)}\sinh^2\be\cr
e^{i\chi}{\sinh2\be\over\sqrt2}&
\cosh 2\be&
e^{-i\chi}{\sinh2\be\over\sqrt2}\cr
-e^{i(\chi-\phi)}\sinh^2\be&
e^{-i\phi}{\sinh 2\be\over\sqrt2}&
e^{-i(\chi+\phi)}\cosh^2\be\cr\end{array}\right)}&\in\SO_0(1,2),\cr
\end{eq}are complex functions on the 2-hyperboloid
$\SO_0(1,2)/\SO(2)\cong \cl Y^2\map\C$ with a triplet representation of
$\SO_0(1,2)$.
However,  there are no finite-di\-men\-sio\-nal  
faithful Hilbert re\-pre\-sen\-tations of noncompact Lie groups.

Relativistic quantum fields
for massive particles involve rectangular transmutators
acted on with $\SL(\C^2)\x\SU(2)$-re\-pre\-sen\-tations.
For example,   the  re\-pre\-sen\-tation $\La=[1|1]$ of the Lorentz group 
 in the $\SO_0(1,3)\x\SO(3)$-re\-pre\-sen\-tation
 on $\C^4\ox V_\io^T$
for a special relativistic vector  field
 and the  tensor  re\-pre\-sen\-tation 
$\La\and\La= [2|0]\pl[0|2]$ 
on $\C^6\ox V_\io^T$  for its field strength,
\begin{eq}{rl}
\bl Z(0)^j
&=\plint_{\cl Y^3} {d^3q\over 2q_0} 
\hskip17.5mm\La({q\over m})^j_a
[\ro u (\rvec q)^a+\ro u^\star (\rvec q)^a],\cr
i\bl F(0)^{kj}
&=\plint_{\cl Y^3} {d^3q\over 2q_0} 
 ~\La({q\over m})^l_0\ep_{lr}^{kj}
\La({q\over m})_a^r

[\ro u (\rvec q)^a-\ro u^\star (\rvec q)^a]\cr

\hbox{with } q_0&=\sqrt{m^2+\rvec q^2},~~
\ep_{lr}^{kj}=\de^k_l\de^j_r-\de^k_r\de^j_l,~ a=1,2,3,~j=0,1,2,3,
\end{eq}are both induced by an $\SO(3)$-re\-pre\-sen\-ta\-tion $[2]$ 
on $V_\io\cong\C^3$ and its dual $V_\io^T$. These spin representations  
 act on the creation and 
annihilation operators
$\ro u(\rvec q)^a,\ro u^\star(\rvec q)^a$ for 
a massive     particle with momentum 
$\rvec q$  and spin 1 directions
$a=1,2,3$. The action of the creation
 operators
 on the Fock ground state
$\rstate 0$ gives dual distributive bases of the special relativistic manifold,
i.e., of the energy-mo\-men\-tum hyperboloid  $\cl Y^3\cong\SO_0(1,3)/\SO(3)$
for mass $m^2$.
The distributive orthogonality is 
 given by the Fock expectation value $\lrangle{~~~}$,
\begin{eq}{l}
\rstate{1;\rvec q,a}=\ro u(\rvec q)^a\rstate 0\in V_\io(\rvec
q),\hskip5mm
\lstate{1;\rvec q,a}=\lstate 0\ro u^\star (\rvec q)^a\in V^T_\io(\rvec
q),\cr
\sprod {1;\rvec p,b}
{1;\rvec q,a}
=\lrangle{\ro u^\star (\rvec p)^{b}\ro u (\rvec q)^a}=
\de^{ab}2\sqrt{m^2+ \rvec q^2}\de(\rvec q-\rvec p).
\cr
\end{eq}The 4- and 6-di\-men\-sio\-nal Lorentz group 
re\-pre\-sen\-tations do not act on Hilbert spaces. That can be seen at
 the transmutators from Lorentz group to rotation group, 
e.g.,  $\{{\rvec q\over m}\mape \La({q\over m})^j_a\}$, which  
are not square integrable $L^2(\cl Y^3)$ on the energy-momentum hyperboloid.

The Lorentz invariant nonnormalizable  measure  of the 3-hyberboloid
in the mo\-men\-tum
pa\-ra\-me\-tri\-zation
cabn be  written as integral with a 
$\cl Y^3$-supported Dirac distribution,
\begin{eq}{rl}
\int_{\cl Y^3} {d^3q\over 2\sqrt{m^2+\rvec q^2}}&=\int d^4q~
\vth(q_0)\de(q^2-m^2).
\end{eq}

The finite-di\-men\-sio\-nal Lorentz group  $\SL(\C^2)$-re\-pre\-sen\-tations
that contain a trivial rotation group $\SU(2)$-re\-pre\-sen\-tation
are $[n|n]$. They  act on vector spaces   $\C^{(1+n)(1+n)}$, $n=0,1,\dots$,
with the spin-representation decomposition:
\begin{eq}{l}
\irrep^{\rm finite}\SO_0(1,3)\ni [n|n]\stackrel{\SO(3)}\cong{\PL_{J=0}^n}[2J]\hbox{ e.g.,  }\left\{\begin{array}{rl}
[0|0]&\cong[0],\cr
[1|1]&\cong[0]\pl[2],\cr
[2|2]&\cong[0]\pl[2]\pl [4].\cr\end{array}\right.
\end{eq}They are 
used for  finite-di\-men\-sio\-nal  $\SO_0(1,3)$-representation 
expansion of complex functions 
on  en\-er\-gy-mo\-men\-tum hyperboloids
 \begin{eq}{rl}
\lstate{[n|n]_0^{j_1\dots j_n}}:&\cl Y^3\map \C
\hbox{ for }n=0,1,2,\dots\cr
\lstate{[n|n]_0^{j_1\dots j_n}}&=\plint_{\cl Y^3}{d^3q\over 2q_0}~ 
{[n|n]_0^{j_1\dots j_n}}(\rvec q)\lstate{ \rvec q}
\hbox{ with }q_0=\sqrt{m^2+\rvec q^2}
\cr
&=\plint d^4q~\vth(q_0)\de(q^2-m^2){[n|n]_0^{j_1\dots j_n}}(q)
\lstate{ q}
\end{eq}and arise as contributions
of the Feynman propagators for trivial translations, e.g., 
for a spin 0 particle in a scalar  field $\pmb\Phi$,
 a spin ${1\over2}$-particle in  a Dirac field $\pmb\Psi$,
  and a   spin 1 particle in a vector field $\bl Z$: 
\begin{eq}{rrl}

\pmb\Phi(0):&\lstate{[0|0]}
&=  \plint d^4q~\vth(q_0)\de(q^2-m^2)
\lstate{ q},\cr
\pmb\Psi (0):&
 \lstate{[0|0]}\pl \ga_j\lstate{[1|1]^j_0}
&=  \plint d^4q~\vth(q_0)\de(q^2-m^2)(\bl 1_4+{\ga_jq^j\over m})
\lstate{ q},\cr
\bl Z^j(0):&\lstate{[2|2]^{jk}_0}
&=  \plint d^4q~\vth(q_0)\de(q^2-m^2)
(-\eta^{kj}+{q^kq^j\over m^2})
\lstate{ q}.\cr
\end{eq}

The spacetime translation dependent fields,  e.g., a massive vector field,
\begin{eq}{l}
\R^4\ni x\mape \bl Z(x)^j,\bl F(x)^{kj}=\ep_{lr}^{kj}{\p^l\over m}\bl Z(x)^r
\end{eq}involve  $e^{iqx}\ro u(\rvec q)^a$ and $e^{-iqx}\ro u^\star (\rvec q)^a$,
which are the translation orbits $\R^4\ni x\mape e^{\pm iqx}\in\U(1)$ 
for a re\-pre\-sen\-tation of the Poincar\'e group $\SO_0(1,3)\sx\R^4$.
This leads to  the spacetime translation representation coefficients
 with $\sprod qx=e^{iqx}$
and  $\sprod xq=e^{-iqx}$ as the
  on-shell part of the Feynman propagator,
\begin{eq}{l}
\sprod{[2|2]^{jk}_0}x+
\sprod x{[2|2]^{jk}_0}
= \int d^4q~\de(q^2-m^2)
(-\eta^{kj}+{q^kq^j\over m^2})
e^{iqx}=\lrangle{\acom{\bl Z^k(y)}{\bl Z^j(x)}}

.\cr
\end{eq}

\end{document}